# Simon-Ando Decomposability and Fitness Landscapes

Max Shpak, Department of Ecology and Evolutionary Biology, Yale University, New Haven, CT 06520-8106, USA

Peter Stadler, Institute of Bioinformatics, University of Leipzig,Kreuzstrasse 7b,D-04103, Leipzig,- Germany

Gunter P.Wagner, Department of Ecology and Evolutionary Biology, Yale University, New Haven, CT 06520-8106, USA

Lee Altenberg, Department of Information and Computer Sciences, University of Hawaii, Manoa HI 96753, USA

## ■ Abstract

In this paper, we investigate fitness landscapes (under point mutation and recombination) from the standpoint of whether the induced evolutionary dynamics have a "fast-slow" time scale associated with the differences in relaxation time between local quasi-equilibria and the global equilibrium. This dynamical behavior has been formally described in the econometrics literature in terms of the spectral properties of the appropriate operator matrices by Simon and Ando (1961), and we use the relations they derive to ask which fitness functions and mutation/recombination operators satisfy these properties. It turns out that quite a wide range of landscapes satisfy the condition (at least trivially) under point mutation given a sufficiently low mutation rate, while the property appears to be difficult to satisfy under genetic recombination. In spite of the fact that Simon-Ando decomposability can be realized over fairly wide range of parameters, it imposes a number of restrictions on which landscape partitionings are possible. For these reasons, the Simon-Ando formalism doesn't appear to be applicable to other forms of decomposition and aggregation of variables that are important in evolutionary systems.

# ■ Introduction

Ever since Wright's (1932) influential characterization of the evolutionary process as a population-level traversal of a genotype or phenotype "landscapes" where evolving populations are attracted to high fitness "peaks" separated from maladaptive "valleys," the metaphor has dominated evolutionary thinking at an intuitive level. However, relatively little has been done to develop a theory of fitness landscape structures as they relate to induced evolutionary processes. This is in part due to the lack of landscape models general enough to be compatible with empirical data. However, perhaps a more fundamental deficiency has been the absence of a theory mapping landscape topology to evolutionary dynamics.

Much of the work that has been done on landscape characterization has dealt with descriptions of statistical features such as autocorrelation over a random walk (Weinberger 1990, Fontana et al 1991, Stadler 1996), or heuristic descriptions of landscape "ruggedness" (Kauffman 1993), measures of neutrality (Ancel and Fontana 2000, Reidys and Stadler 2001) and topological descriptions of connectivity (Gavrilets and Gravner 1997). These landscape characteristics were chosen with the intuition that they should somehow correlate with evolutionary dynamics. For example, a "rugged" landscape can be thought of as having more local attractors (or rather, quasi-equilibria in a connected landscape) than a "smooth" landscape, while connectivity should relate to the ability of the evolving population to move between local peaks given potentially infinite time. However, most of these heuristic characterizations did not make an explicit link to the dynamical system properties correlated with the given measure.

In what follows, we investigate the possibility of characterizing fitness landscapes according to whether the state description induced by a particular landscape (or, rather, its corresponding mutation-selection operator) is decomposable and aggregable, or at least are decomposable and aggregable to a close approximation at the appropriate scale. We will begin by briefly defining what we mean by "decomposability" and "aggregability," while referring the reader to a more detailed treatment of these subjects in Shpak et al (2003).

In the broadest sense, aggregability of a dynamical system means that there exist macrostate variables (which are themselves functions of some subset of the microstate variables) which allow for a dynamically sufficient description of the system with fewer state variables than in the original dynamical system. The ability to collapse subsets of state variables into an aggregate variable on each equivalence class reflects symmetries inherent in the selection and transmission processes specified by landscape topology and the mutation operators. From a purely computational point of view it is valuable to identify such symmetries, because they render otherwise computationally expensive analyses tractable by reduction of variables. In other words, from a practical standpoint, the advantage of a reduced state-space description are obvious: any reduction in the state space necessary for a dynamically sufficient description substantially simplifies both numerical and analytical treatments of the process in question.

In addition to the obvious significance of aggregation of variables from the perspective of computational complexity, the dynamical symmetries induced by a particular decomposable landscape shed light on the properties of the evolutionary process. The fitness landscape symmetries that make aggregation of variables possible may relate with hierarchy and modularity in the genotype-fitness map or other organizational constraints. Indeed, the general concept of aggregation has been widely applied to a range of biological models and problems, from classical notions of modularity in organismal design to modern notions of hierarchy in genetic architecture (Frenken et al 2000, Simon 1972 and 2000, unpublished).

System decomposability is often related to aggregation of variables but is in principle an independent property. It simply means that there exist (usually) non-intersecting subsets of variables which interact among themselves in some way that distinguishes their interaction from any other subset of variables. In other words, it implies the existence of self-contained "modules" of state variables in a dynamical system that are in some way independent or quasi-independent of the other microstate variables. Often these modules have associated aggregation rules that allow them to be characterized as "emergent" macrostate variables, but this need not be the case (Shpak et al, 2003).

In more formal terms, consider a discrete dynamical system (what follows is trivially generalizable to continuous systems) specified by a linear operator **A**, such that x(t+1)=**A**x(t). For convenience, we will denote x(t) as x and x(t+1) as y. We define aggregativity to mean that given a dynamical system with $(x_1...x_n)$ state variables, one can group subsets of the microstate variables into m<n macrostates $X_1=f_1(x_1...x_n)...X_m=f_m(x_1...x_n)$, where in the simplest scenario f is simply a linear combination of microstate values. These macrostates in turn must constitute a dynamically sufficient description, i.e. so that there exists an operator $\hat{A}$ such that X(t+1)=$\hat{A}$X(t).

For a linear system specified by transition operator **A**, we require that there exists an aggregation operator **Q** such that there is an aggregate variable description Y=$\hat{A}$X, where Y=**Q**y and X=**Q**x. If y and x are state vectors of length n, for the aggregation **Q** to be non-trivial, it must be an mxn matrix (with m<n), and with $\hat{A}$ an mxm matrix. Such an aggregation of variables achieves a reduction of the state space dimension necessary for a dynamically sufficient description. **Q** is defined such that if $Q_{Ij}>0$, $x_j$ is a member of the Ith aggregate class (the indices I,j refer to the element in the Ith row and jth column of the matrix **Q**, the use of upper and lower-case letters is simply to indicate that the nonzero indices j are within the Ith subset).

It was shown by Shpak et al that whenever such an aggregation exists, the aggregate dynamics operator $\hat{A}$ has the form

$$(1.1) \quad \hat{A} = QAQ^T \left(QQ^T\right)^{-1}$$

It was also proven that even for systems which are not exactly aggregable, (1.1) gives the best approximation for the aggregate dynamics of X given the aggregation rule **Q**.

In the special case where the subsets defining the aggregate classes are mutually exclusive, the dynamical system is said to be decomposable as well as aggregable. The set of such subclasses is refered to as a partitioning, and the aggregation of variables on the partition have a **Q** matrix structure such that if $Q_{Ij}>0$ then $Q_{Kj}=0$ for all other K≠I. This means that the dynamical system in question is in some way "modular," i.e. variables contributing to a given macrostate variable do not contribute to others. Together, these rules specify an aggregation of variables which is also a decomposition. As discussed in the first section of Shpak et al (2003), decomposability and aggregativity are in principle independent of one another.

Contra the idealized representations of aggregation and decomposability presented above, most natural systems have a certain amount of communication (at least indirect) between variables across partitions. Indeed, when complete decomposability holds, the system is rather uninteresting in that it consists of non-interacting, self-contained subcomponents. One would expect a much more common scenario to be one where a dynamical system is said to be nearly decomposable if the aggregation via **Q** is dynamically sufficient as an approximation within some arbitrarily small error parameter $\epsilon \ll 1$, such that the linear operator **A** can be written as **A**=$A^*$+$\epsilon$**C**, where $A^*$ is completely decomposable and **C** is an arbitrary n dimensional square matrix.

One such class of nearly decomposable dynamical systems occurs when the macrostate variables are specified as subsets which are nearly independent from the remaining state variables and exhibit local quasi-equilibria, at least over a certain time scale. We propose that such models may be of particular interest to population genetics from the standpoint of peak transitions in evolution. Such systems have a dynamic that is characterized by short-term clustering about local equilibria and long-term transitions towards the global equilibrium. It is hypothesized that the subsets of variables which exhibit local quasi-equilibria correspond to local fitness optima and their mutational neighbors (with the long time scale equilibrium being the mutation-selection balance over the entire landscape), and that such fitness landscapes induce "fast-slow" (e.g. Guckenheimer and Holmes 1981, Strogatz 1994) behavior.

Examples of nearly decomposable and aggregable natural systems include many well-known models in thermodynamics and statistical physics, which use macrostate variables such as temperature, pressure, and entropy as state variables. This permits a dynamically sufficient descriptions of particle ensembles which would be impossible to describe by tracking the microstates of the innumerable gas particles. In biology, aggregation of microstates into macrostates is (at least implicitly) the foundation for quantitative genetics, where the phenotype macrostates serve as (often approximate) state variables, and the corresponding microstates are the contribution to phenotype or mutational variance made by each individual locus (e.g. Bulmer 1970, Turelli and Barton 1994).

The purpose of this study is to demonstrate the applicability of aggregation methods in describing mutation-selection dynamics on model fitness landscapes, particularly the use of decomposability criteria to identify fitness landscapes which induce fast-slow evolutionary dynamics. It is hypothesized that those landscapes which are characterized by local quasi-equlibria are decomposable and aggregable into subsets of genotypes defining each quasiequilibrium. We begin with a general analysis of decomposability of linear dynamical systems, because of its simplicity and the fact that mutation-selection operators for haploid genotypes can be readily linearized.

## ■ Simon-Ando Decomposability and Fast-Slow Dynamics

Aggregativity and decomposability in what we will refer to below as "Simon-Ando" systems is only approximate, and as such the aggregate representations are only accurate over durations determined by the system dynamics (insofar as the near-aggregativity arises due to a time scale decoupling between strong within-partition dynamics and weak cross-partition dynamics). We begin with a summary of Simon and Ando's results on fast-slow dynamical behavior as far as is necessary for understanding their relevance to what follows, while referring the reader to their papers and to Courtois (1977) for a more detailed discussion.

Consider a system of linear difference equations x(t+1)=**A**x(t) where **A** is an nxn transition matrix. Suppose that **A** can be rewritten as

(3.0) $\quad \mathbf{A} = A^* + \epsilon \mathbf{C}$

with $A^*$ is a block diagonal matrix of the form:

$$\begin{pmatrix} A_1^* & 0 & 0 & 0 & 0 \\ 0 & \ddots & 0 & 0 & 0 \\ 0 & 0 & A_I^* & 0 & 0 \\ 0 & 0 & 0 & \ddots & 0 \\ 0 & 0 & 0 & 0 & A_M^* \end{pmatrix}$$

Each element $A_I^*$ is a submatrix of dimension $n_I$ (i.e. the number of elements in the partition class $C_I$) such that $n=\sum_{i=1}^{k} n_I$, while the corresponding partitions of the state vector $x^*$ on which the submatrices operate are denoted as:

$$x^*(t) = \{x_i^*(t)\} = \{[x_{i_1}^*(t)],...,[x_{i_I}^*(t)],...,[x_{i_M}^*(t)]\}$$

where $x_{i_I}^*(t)$ is the subset of components of $\{x_i^*(t)\}$ where the indices given by $x_{i_I}^*(t)=x_i(t)$ for $x_i \in C_I$, with the lowest valued index with the partition being $i=\sum_{J=1}^{I-1} n_J + j$. In other words, the state vector $x_i^*(t)$ is divided into I=1...M subvectors of length $n_I$ corresponding to the dimensions of $A_I^*$. The values of $A_J^*$ for J≠I have no effect on the behavior of $x_i^* \in C_I$. In this scenario, the system's behavior is fully specified by:

(3.1)   $x_i^*(t) = A_I^{*t} x_i^*(0)$

Due to the fact that there are non-communicating blocks in $A^*$, the dynamical system determined by this matrix satisfies the requirements of complete partitionability: namely, that the system dynamics can be fully specified by applying transformation rules to each partition independent of the other.

In the system specified by $A=A^*+\epsilon C$, however, there is weak communication (scaling with $\epsilon \ll 1$) across states, because **C** has nonzero off-diagonal (defined relative to $A^*$) entries. The existence of strong communication (approximated by the transition rates of $A_I^*$) for variables within partitions I,I and correspondingly weaker communication for variables across classes I and J (I≠J) leads to a fast-slow dynamic. Furthermore, as will be argued below, it allows one to (approximately) construct an aggregate dynamical system where the macrostate variables $X_I$ are functions of the elements of the Ith strongly-communicating partition class of vertices $[x_{i_I}(t)]$.

At an intuitive level, approximate aggregation is possible because the distribution of variables $[x_{i_I}(t)]$ corresponding to partition class $C_I$ tends towards a quasi-equilibrium very rapidly compared to the time required for the system as a whole to reach the global equilibrium. Consequently, for some intermediate time period, the subvectors $[x^*_{i_I}(t)]$ for each partition class will be in quasi-equilibrium determined by the corresponding transition operator $A^*_I$, while an aggregate transition matrix of between-class transition rates can adequately approximate the transition dynamics between classes.

To see that this is indeed the case, consider the eigenvalues of each submatrix $A^*_I$. Assuming that they are distinct, the $n_I$ eigenvalues can be written in descending order so that $\lambda^*_{1_I} > \lambda^*_{2_I} > ... > \lambda^*_{n_{I_I}}$. Furthermore, the constituent blocks $A^*_I$ of matrix $A^*$ can be permuted so that the leading eigenvalues of each submatrix are arranged in the order $\lambda^*_{1_1} > \lambda^*_{1_I} > ... > \lambda^*_{1_M}$ for the m submatrices (where $\lambda^*_{i_I}$ denotes the ith eigenvalue of the Ith block).

In order to describe the decoupling of dynamics between $\mathbf{A}$ and the $A^*$ approximation, define a value $\delta^*$ such that:

(3.2a) $$\min_{i \neq j} |\lambda^*_{i_I} - \lambda^*_{j_I}| < \delta^*$$

The difference in eigenvalue magnitudes in the components of $A^*$ are next compared to those of $\mathbf{A}$, so that for any positive real number $\delta < \delta^*$ there is a sufficiently small $\epsilon > 0$ (as in eq. 2.0) such that for all $i \in C_I$

(3.2b) $$|\lambda_{i_I} - \lambda^*_{i_I}| < \delta$$

The differences in magnitude between $\delta$ and $\delta^*$ determine the characteristic time scale differences between the fast and slow dynamics. The fast system dynamics of $x_{i_I}(t)$ will be driven by the eigensystem of the Ith subcomponent, hence being nearly identical to the behavior of $x^*_{i_I}$. In the short run (for a time span which scales exponentially with the magnitude of $\delta$), the system behaves as a completely decomposable system specified by $A^*$. Given sufficient time, the subsystems settle into quasiequilibria given by the leading eigenvalue $\lambda_{1_I}$ and the associated eigenvector of each submatrix, with the remaining (up to) $n_I-1$ eigendirections being residual. The system x(t) as a whole consists of M quasiequilibria determined by their eigenvectors $v_{1_1}...v_{1_M}$.

The system then becomes driven primirily by communication across quasi-equilibrium subcomponents, tending ultimately towards the eigendirection corresponding to the leading eigenvalue $\lambda_1$ of $A$. This can be seen by writing a spectral decomposition of the transition matrix $A$ and its powers, i.e.

$$(3.3) \quad A^t = \lambda_{1_1}^t Z_{1_1} + \sum_{I=2}^{m} \lambda_{1_I}^t Z_{1_I} + \sum_{I=1}^{m} \sum_{i=2}^{n_I} \lambda_{i_I}^t Z_{i_I}$$

Where $Z$ is the matrix of the product of left and right eigenvectors of $A$, i.e. $Z=v\tilde{v}$ for $\lambda v = Av$ and $\lambda\tilde{v}=\tilde{v}A$, given an appropriate choice of normalization constants. $Z$ satisfies the relation $AZ=Z\Lambda$ (with $\Lambda$ a diagonal matrix of $A$'s eigenvalues ordered by descending absolute value).

The corresponding spectral expansion of $A^{*t}$ (the completely aggregable approximation to $A^t$) is

$$(3.3\,b) \quad A^{*t} = \sum_{I=1}^{m} \lambda_{1_I}^{*t} Z_{1_I}^* + \sum_{I=1}^{m} \sum_{i=2}^{n(I)} \lambda_{i_I}^{*t} Z_{i_I}^*$$

Where $\lambda^*$, $Z_I^*$ are the eigenvalues and eigenvector products of $A^*$. From the properties of near-decomposability, $\lambda^*_{iJ} \rightarrow \lambda_{iJ}$ and $Z^*_{iJ} \rightarrow Z_{iJ}$ if in (3.0) $\epsilon \rightarrow 0$ (Simon and Ando 1961, pgs. 118-21). In the special case of stochastic matrix $A$ and $A^*$ (with every submatrix of the latter also stochastic), we also have the condition $\lambda_{1_1} = \lambda^*_{1_1} = \lambda^*_{1_J} = 1$.

For the above spectral decomposition, the jth component of the state vector $x_{j_J}(t) = A^t x_{j_J}(0)$ can be written as

$$(3.4) \quad x_{j_J}(t) = \sum_{i=1}^{n} Z_{ij} \lambda_i^t y_i(0)$$

where $y(t) = Z^{-1} x(t)$. By expanding and rearranging terms, is

$$(3.4a) \quad x_{j_J}(t) =$$

$$\lambda_{1_1}^t Z_{1_1 j_J} y_{1_1}(0) + \lambda_{1_J}^t Z_{1_J j_J} y_{1_1}(0) + \sum_{I \neq J, I=2}^{m} \lambda_{1_I}^t Z_{1_I j_J} y_{1_I}(0) +$$

$$\sum_{i_J=2}^{n_J} \lambda_{i_J}^t Z_{i_J j_J} y_{i_J}(0) + \sum_{I \neq J, I=1}^{m} \sum_{i_I=2}^{n_I} \lambda_{i_I}^t Z_{i_I j_J} y_{i_I}(0)$$

In order to satisfy the properties of near decomposability, $Z_{i_I j_J}$ must be orders of magnitude smaller than eigenvector components $Z_{i_I j_I}$ due to the significantly stronger within vs. between component couplings. Similarly, the completely aggregable approximation contains no cross-component terms (I,J) because the block-diagonal matrix has no cross-term communication. From the fact that $Z^*_{i_I j_J} = 0$, we write:

$$(3.4b) \quad x^*_{j_J}(t) = \lambda_{1_J}^{*t} Z^*_{1_J j_J} y_{1_1}(0) + \sum_{i_J=2}^{n_J} \lambda_{i_J}^{*t} Z^*_{i_J j_J} y_{i_J}(0)$$

The short term behavior induced by $A^t$ (2.3) can be broken into intervals $T_1 < T_2 < T_3$ such that in the first two intervals (2.3b, 2.4b) are good approximations. That such a time-scaling exists follows from the Simon-Ando theorems, (Thms. 4.1 and 4.2 in Simon and Ando 1961, which we repeat here without proof):

**Theorem 2.1:**

Given Z and $Z^*$ defined as in (3.3), for an arbitrarily small scalar quantity $\xi > 0$, there exists a sufficiently small $\epsilon < \epsilon_\xi$ (as in 2.0) such that

$$\max_{k,l} | Z_{kl}(i_I) - Z^*_{kl}(i_I) | < \xi$$
$$\text{for } 2 \leq i \leq n(I), \ 1 \leq I \leq N, \ 1 \leq k, l \leq n$$

Theorem (2.1) allows one to precisely bound the difference between the eigenvector components of the completely and nearly aggregable systems, and to combine this information with the ranking of eigenvalue magnitude in making predictions about the qualitative short-term versus long-term behavior of the system. Defining $\xi_{kl} = Z_{kl}(i_I) - Z^*_{kl}(i_I)$, following Simon and Ando we write

$$(3.5) \quad u_{ij} = \frac{\xi_{ij}}{\xi}$$

and

$$(3.6) \quad Z = Z^* + \xi U$$

This notation (together with the fact that for cross-component terms IJ the components $Z_{IJ} = 0$) allows us to rewrite the five terms of (2.4a) as

$$(3.7\,a) \quad x_{j_J}(t) = \xi \lambda_{1_1}^t u_{1_1 j_J} Y_{1_1}(0) +$$

$$\lambda_{1_J}^t z_{1_J j_J} Y_{1_1}(0) + \xi \sum_{I \neq J, I=2}^{m} \lambda_{1_I}^t u_{1_I j_J} Y_{1_I}(0) +$$

$$\sum_{i_J=2}^{n_J} \lambda_{i_J}^t z_{i_J j_J} Y_{i_J}(0) + \xi \sum_{I \neq j, I=1}^{m} \sum_{i_I=2}^{n_I} \lambda_{i_I}^t z_{i_I j_J} Y_{i_I}(0)$$

which we rewrite using the shorthand of Simon and Ando's equation (4.21),

$$(3.7\,b) \quad x_{j_J}(t) = \xi S_j^{(1)} + S_j^{(2)} + \xi S_j^{(3)} + S_j^{(4)} + \xi S_j^{(5)}$$

where the S terms are shorthand for the each of the respective terms in (3.7a).

Note that the first term (the leading eigendirection with corresponding leading eigenvalue $\lambda_{1_1}$) scales as $\xi \ll 1$ because it involves I,J cross-component interaction. Consistent with a qualitative account of system dynamics, the short term behavior of $x_{j_J}(t)$ is largely dominated by $S^{(2)}$ and $S^{(4)}$ (the only terms which appear in the completely decoupled system described by (2.4b)). For larger values of t, however, eventually $S^{(1)}$ comes to dominate as $\lambda_{1_J}^t \ll \lambda_{1_1}^t$ (and other terms) as $t \to \infty$ (and the global equilibrium is attained). The same applies for determining the time scale at which quasi-equilibria are attained within each Ith subcomponent ($S^{(2)}$ vs. $S^{(4)}$).

We can state this more precisely by delimiting time intervals over which some subsets of terms in (2.7) dominates over the others. This is made explicit in Theorem (4.2) in Simon and Ando,

**Theorem 2.2a: (Within-Component Quasi-Equilibria)**

Because $\lambda_{1_J} > \lambda_{j_J}$ for $j \geq 2$, there exists a $T_0$ such that for $t > T_o$

$$(3.8) \quad \frac{S_j^{(4)}}{S_j^{(2)}} < \eta_0$$

While for $t > T_0$ and some real number bound $\eta$, since from theorem 2.1 it follows that there exists a $\xi$ for $\epsilon$ so that for $T_0 < t < T_1$

$$(3.9) \quad \frac{\xi \left( S_j^{(1)} + S_j^{(3)} + S_j^{(5)} \right)}{S_j^{(2)} + S_j^{(4)}} < \eta_1$$

Theorem 2.2a means that for a sufficiently small elapsed time, the dynamics of x(t) is dominated by within-component interactions. The first inequality (3.8) describes the short-term within-component behavior before each partition class approaches its local quasiequilibrium. When $S^2$ dominates, the component is described as being in quasi-equilibrium. The second inequality simply states that for time less than $T_1$, the cross component terms of order $\xi$ are negligible.

The next set of inequalities describe the behavior of cross-component (I,J) interactions. The first establishes a time limit at which cross-component interactions first become significant, the second the waiting time until the (global) leading eigendirection dominates:

**Theorem 2.2b: (Cross-Component Equilibrium)**

Given an $\epsilon$ which gives the bounds in Theorem 2.2a, given a bound $\eta_2$ there exist time $T_2 > T_1$ such that for $T_2 < t$,

$$(3.10) \quad \frac{S_j^{(4)} + \xi S_j^{(5)}}{\xi S_j^{(1)} + S_j^{(2)} + \xi S_j^{(3)}} < \eta_2$$

for some value $\eta_3$, there exists a number $T_3 > T_2$ so that for $t > T_3$

$$(3.11) \qquad \frac{S_j^{(2)} + \xi S_j^{(3)} + S_j^{(4)} + \xi S_j^{(5)}}{\xi S_j^{(1)}} < \eta_3$$

The inequality (3.10) states that beyond time $T_2$, the cross-component terms start to dominate over the within-component terms, while (3.11) suggests that given sufficient time $t \gg T_3$, the leaning eigendirection (the direction of the global equilibrium) dominates all other terms.

The inequalities in Theorem 2 divide the fast-slow dynamics of x(t) into four stages. Consider an initial state vector x(0), in particularly the Ith component of the probability distribution $x_I(0)$.

Fort the entire time interval $t<T_1$, the behavior of $x_I(t)$ is closely approximated as $x^*_I(t)$, i.e. by the equations (3.4b). In the time interval $0<t<T_0$, the distribution $x_I(t)$ is determined by the various eigendirections associated with $A_I$, while during the time interval $T_0<t<T_1$, the leading eigenvector of $A_I$ dominates, and $x_I(t)$ is said to be in a state of local quasi-equilibrium.

When $T_1<t<T_2$, cross component interactions with partition classes $J \neq I$ become significant. Since prior to time $T_2$ the distribution $x_I(t)$ is at a quasi-equilibrium distribution closely approximated by the eigenvector $Z_{1_I}$ (and the same is the true for all subcomponents J), the dynamics during the time intervals following $T_1$ can be approximated by assuming that the Ith component is in the equilibrium associated with $A^*_I$, weighted by what initial proportion of the distribution was in the Ith component.

This is the property that allows for aggregation of variables approximations to be reasonably accurate representations for the dynamics of x(t). By treating the within-component distributions at quasiequilibrium as essentially static over the short term, each subcomponent distribution $x_I$ can be treated as an aggregate variable $X_I$, and rather than tracking the interactions of individual elements i,j of the Ith and Jth component, the dynamics for $T_1<t<T_2$ can be described by an aggregate matrix $\hat{A}$ that describes net cross-component communication (as described in the following section).

During the time interval $T_1 < t < T_2$, the cross-component eigenvectors begin to dominate over the within-component eigenvectors, while for $T_2 < t$, the leading eigenvector of **A** dominates the distribution as it tends towards its global stable equilibrium (assuming that such exists). True fast-slow behavior occurs when $T_2$ is substantially (perhaps orders of magnitude) larger than $T_0$ by allowing a time scale decoupling between local quasiequilibria and global equilibration.

# ■ Aggregation of Variables and Decomposition in Simon-Ando Systems

In a nearly decomposable Simon-Ando system at time $T_1 < t < T_2$, the distribution in each component $x_I(t)$ is in quasi-equilibrium with nearly stationary within-partition class dynamics. Since for larger values of t cross-component (I,J) terms begin to dominate the system dynamics, the quasiequilibrium distributions can be treated as macrostate variables with a time evolution determined by the sum total of transition rates across classes.

Given a transition matrix **A** which induces Simon-Ando dynamics, we construct a description of transition between macrostates (i.e. exchange rate between the K diagonal subcomponents in $A^*$) by summing the indices $A_{ij}$ across classes iϵI and jϵJ. Simon and Ando (1961) and Courtois (1977) derive an aggregate dynamics operator $\hat{A}$ for a stochastic matrix **A** (their results apply readily to non-stochastic operators, apart from the leading block-eigenvalues not being equal to one) by arguing that in time interval $T_1 < t < T_2$, the normalized Ith component of $x_I(t)$ is approximately equal to the leading eigenvector (stationary distribution) of $A_I^*$,

$$(3.12) \quad v^*_{iI}(1_I) \simeq \frac{x_{iI}}{\sum_{i \in I} x_{iI}}$$

We next construct a transition matrix which describes the dynamics of the transition probability between states J and I. In terms of conditional probabilities (and by abuse of notation where X=J denotes a transition to the Jth state), we have

$$P(X_{t+1} = I \mid X_t = J) = \frac{P(X_{t+1} = I, X_t = J)}{P(X_t = J)}$$

with $P(X_t = J) = \sum_{j \in J} x_{jJ}$ and $P(X_{t+1} = I, X_t = J)$ is the probability of leaving any state i∊I entering any state j∊J, which is the sum of $A_{ij}$ transition probabilities weighted by the probability of being in the ith state, i.e. $\sum_{i \in I} \sum_{j \in J} x_{jJ} \mathbf{A}_{ij}$. Therefore,

$$(3.13) \quad P(X_{t+1} = I \mid X_t = J) = \frac{1}{\sum_{j \in J} x_{jJ}} \sum_{i \in I} \sum_{j \in J} \mathbf{A}_{ij} x_{jJ}$$

with identical results (apart from a normalization constant) if one interprets the coefficients of **A** as rate terms rather than transition probabilities. From (2.9), we can approximate this expression as

$$(3.13\,b) \quad P(X_{t+1} = I \mid X_t = J) \simeq \sum_{i \in I} \sum_{j \in J} \mathbf{A}_{ij} v^*_{jJ}(1_J)$$

For the completely decomposable system $A^*$, by (3.13b) the transition rates are simply the elements of a diagonal matrix of leading right eigenvalues for each block, $\boldsymbol{P}_{II}=\lambda(1_I)$, $\boldsymbol{P}_{IJ}=0$ for I≠J, with the former coefficient equal to unity for block-stochastic matrices.

This construction of aggregate dynamics implicitly assumes an aggregation operator **Q** which is a matrix of characteristic vectors, as in this formulation $X_I = \mathbf{Q}x = \sum_{i \in I} x_{iI}$. However, a derivation of $\hat{A}_{IJ}$ using the matrix of characteristic vectors according to (1.2) gives us something of the form (3.5), which is not equal to (3.13).

Alternatively, one could chose **Q** to be the matrix with rows $q_I = v^*_I(1_I)$, the leading eigenvectors of the block-diagonal submatrices of $A^*$ to determine whether this choice of aggregation operator is consistent with (1.1). In this case, $X_I$ would have to be normalized to produce a distribution as $\sum_{i \in I} v^*_I(1_I) x_{iI}$ is not necessarily normal given x, $v_I^*$ normal. Furthermore, the resulting derivation of $\hat{A}$

$$\hat{A}_{IJ} = \sum_{i \in I} \sum_{j \in J} v^*_{iI} A_{ij} \frac{v^*_{jJ}}{\sum_{k \in J} v^*_{kJ}{}^2}$$

is only equivalent to (3.13) in the special case where each block submatrix of $A^*$ has as its leading eigenvector a uniform stationary distribution.

This suggests that the standard expression (3.13) for the aggregate transition matrix in Simon-Ando systems does not generally satisfy the mean-square minimum of Theorem 1.1, nor is the $\hat{A}$ used in Simon-Ando systems consistent with any choice of aggregation operators given (1.1). As discussed above, this does not mean that the expression for $\hat{A}$ consistent with (1.1) should be preferred to a different aggregate matrix (3.10), as Theorem 1.1 only offers a heuristic for the construction of $\hat{A}$ rather than a rigorous statement about the best approximation for aggregate dynamics. In fact, much of the numerical work related to Simon-Ando systems (Courtois 1977, Meyer 1989, Kafeety et al 1996, Deuflhard et al 2000) suggests that for strong-time decoupled systems with a large separation $T_2 - T_1$, the aggregation approximation for the dynamics of X is quite robust.

One of the differences between (3.13) and (1.1) is that the latter gives the least-square minimum estimate of $\|\mathbf{QA}x - \mathbf{AQ}x\|$ for x(t) over all times t, while (3.13) should give the best estimates for $X = \mathbf{Q}x$ for time intervals $t > T_1$. Some of our numerical results in the sections below

suggest that for certain parameter landscapes (presumably, those where Simon-Ando decomposability might be a poor approximation), (1.1) gives reasonable estimates of aggregate dynamics where (3.13) fails. Conversely, (3.13) gives good estimates of the dynamics of **Q**x and the stationary distributions of the transition operator A in cases (presumably those which exhibit Simon-Ando dynamics) where (1.1) gives completely misleading estimates.

The relationships above suggest the time scales over which a Simon-Ando system is (to a close approximation) aggregable and decomposable, as the two properties are not necessarily congruent. During the time interval t<$T_0$, system dynamics are dominated by within-partition processes, and thus the system dynamics can be described to a close approximation in terms of the within-partition dynamics of $x_I$ acted on by their respective $A_I^*$.

For the time interval $T_0$<t<$T_1$, because the within-partition dynamics approximated by $A^*$ still dominate, the system remains decomposable. However, the distributions within each partition are closely approximated by the eigenvectors $v_{1_I}^*$, hence in this interval the system is aggregable. In contrast, over the time interval t<$T_0$ the dynamical system was decomposable but not aggregable.

For $T_2$<t, cross-partition communication becomes dynamically significant, thus $A^*$ is no longer a good approximation for the system as a whole. The exchange rates across partitions have to be weighted by their quasistationary within-component distributions, so that the dynamical system in this time interval remains nearly aggregable in spite of the breakdown of decomposability.

The different behaviors over the defined time intervals illustrate the fact that for Simon-Ando systems where aggregation and decomposition are only approximate and time-dependent, the two properties are actually decoupled from one another. This is in contrast to the situation one encounters in dynamical operators that can be aggregated according to exact equitable partitions (Stadler and Tinhofer 1999, Shpak et al 2003). Those systems are exactly decomposable and aggregable over all time scales.

## ■ Mutation-Selection and Fitness Landscapes

A fitness landscape (V,$\xi$,f) consists of a vertex set of genotypes V, a transmission rule $\xi$ that defines neighborhood relationship and/or a distance metric between vertices (genotypes) i,j, and a real-valued function f:V→$\mathbb{R}$ which assigns a fitness value to each genotype (Weinberger 1990, Jones 1992, Culberson 1992, Stadler 1994). In genetic systems, the relevant transmission rules are those that assign transition probabilities (edge weights) $T_{ij}$ between genotypes i and j for mutation and recombination processes. We will first treat the simplest case, the fitness landscape specified by a mutation-selection process under single point mutation.

Consider a haploid asexual population, with k possible genotypic states. These can be interpreted as either k alleles at a single locus, or alternatively, k alternative multilocus genotypes. The state vector x(t) describes frequencies of ith genotype $x_i$(t), where $\sum_{i=1}^{k} x_i = 1$. To each genotype one assigns fitness value $w_i$, and define the mean population fitness as $\overline{W} = \sum_{i=1}^{k} x_i w_i$.

For the sake of simplicity, we assume that reproduction, mutation, and selection occur in non-overlapping generations. The probability that a j individual has an i offspring is given by the per-generation mutation rate $\mu_{ij}$. This lets us define a row-stochastic mutation matrix **M** with entries:

$$(4.0) \quad M_{ii} = 1 - \sum_{i=1}^{N} \mu_{ji}; \quad M_{ij} = \mu_{ij} \text{ for } i \neq j$$

Iteration of x(t) across a single generation is defined by an expression of the form:

$$(5.1) \quad x_i(t+1) = x_i(t) \frac{w_i}{\bar{W}} + \frac{1}{\bar{W}} \sum_{j=1}^{k} (\mu_{ij} w_j x_j - \mu_{ji} w_i x_i)$$

(see Crow and Kimura 1970, Ewens 1979, Burger 1998, Burger 2001, etc.).

The mutation-selection dynamics can be fully specified in the form of a matrix **A** operating on x, with

$$(5.2) \quad A_{ij} = \mu_{ij} w_j \text{ if } i \neq j$$

$$A_{ii} = \left(1 - \sum_{l=1}^{k} \mu_{il}\right) w_i$$

The mutation-selection matrix **A**=**WM**, where **M** is the mutation matrix (5.0) and **W** is a diagonal matrix of genotype fitness values $W_{ii}$=w, $W_{ij}$=0.

By definition $\bar{W} = \sum_{i=1}^{k} (Ax(t))_i$, thus one can write (5.2) as $x_i(t+1) = \frac{1}{\bar{W}} Ax_i(t)$. In its general form, one can describe the dynamics as:

$$(5.3) \quad x(t) = A^t x(0) / \sum_{i=1}^{k} \left(A^t x(0)\right)_i$$

The same equations will describe evolution of a diploid population, if one defines $w_i$ as a marginal fitness of genotype i, i.e. $w_i = \sum_{j=1}^{k} w_{ij} x_j$ and $\bar{W} = \sum_{ij=1}^{k} w_{ij} x_i x_j = \sum_{i=1}^{k} x_i w_i$.

It can be seen that eq. (5.3) is nonlinear due to the occurrence of terms containing $x_i$ in $\bar{W}$. The characterization of a number of dynamical system properties is more straightforward if the equations can be linearized and expressed in matrix form. The decomposability criteria outlined in the first section only holds for linear systems.

A linearization of the coupled macromolecule synthesis models of Eugene (1971) by Jones et al (1976) and in Thompson and MacBride (1974). These are formally equivalent to the Crow-Kimura mutation-selection models, thus the treatment here is a transformation which is equivalent to that of Jones et al. Consider the transformed variable:

$$(5.4) \quad x(t) = y(t) \exp\left[-\int_0^t \frac{1}{\bar{W}} dt'\right]$$

$$\text{with } \bar{W} = \sum_{i=1}^{k} x_i(t') w_i$$

Substituting (5.4) into (5.1b) we obtain a linear difference equation. Note that this can be written in matrix form similar to (5.2) but without the nonlinear normalization term $\bar{W}$. This linearization is only valid for haploid genotypes (Thompson and MacBride 1974) because the diploid to haploid (gametic) transmission process is quadratic.

The linearization of the mutation-selection equations essentially changes the description of relative genotype frequencies to one of absolute frequencies (Baake et al 1997, Hermisson et al 2001) in an exponentially growing population rather than relative frequencies and requires the use of (absolute) Malthusian rather than Wrightean fitness parameters. One can transform back to the Wrightean relative frequency and fitness description with no information loss (by normalizing the absolute frequency vector), and the mutation-selection equations for haploids can be treated as a linear system to which the above definitions of decomposability are applicable.

In an absolute-frequencies based representation, the mutation-selection process can be fully specified as **A**=**WM**, i.e. as the product of a diagonal matrix of genotype fitness values and a stochastic mutation matrix (this is where the discussions in previous sections on the conservation of stochasticity in $\hat{A}$ becomes relevant). Furthermore, when absolute frequencies are used and the nonlinear $\overline{W}$ term is removed, the linear operator **A** in (5.2) gives us the state equation for gene frequency change from one generation to the next.

The goal in constructing an aggregation-based description of mutation-selection processes is, given (5.1), to be able to construct either as an exact or approximate description

$$(5.5) \quad X_I(t+1) = X_I(t) W_I + \sum_{J=1}^{k} M_{IJ} W_J X_J - M_{JI} W_I X_I = (\hat{\mathbf{A}} X(t))_I$$

where $X_I$ is the aggregate frequency of the Ith class of genotypes, $M_{IJ}$ is some measure of net mutation rate from the Ith to the Jth class (usually as some weighted sum of cross-term mutation rates $\sum_{i \in I, j \in J} c_{ij} \mu_{ij}$, and $W_I$ is some aggregate fitness value on the Ith aggregate class, i.e. $W_I = f(x_{I_1} ... x_{I_{n_I}})$ where $x_{I_i} \in C_I$.

# ■ Landscapes on Hamming Graphs and Simon-Ando Decomposability

We now turn our attention to the question of whether Simon-Ando decomposability is applicable to a wide class of fitness landscapes. As stated in the introduction, one's intuition about landscape topography suggests that Simon-Ando like behavior might be fairly generic in multi-peaked landscapes where the peaks and surrounding high-fitness genotypes are separated by broad, low-fitness "valleys." We hypothesized that genotype frequency distributions would reach quasi-equilibria in the neighborhoods of local optima in the short run while slowly moving across the valleys towards the global mutation-selection equilibrium over a longer time scale.

The qualitative behavior and some of the mathematical structure associated with Simon-Ando dynamics has been observed in a variety of evolutionary models. The "epochal evolution" of the "Royal Road Genetic Algorithm" (van Nimwegen et al 1997) exhibits local quasi-steady states associated with the fixation of a subset of loci with a certain fitness effect, punctuated by stepwise transitions between these steady states in attaining the global optimum. Furthermore, the block-diagonal structure of fitness landscapes and its associated eigenstructure has been used by Schuster and Swetina (1988) and Wilke (2001a,b) to derive equilibrium distributions and determine global optima by identifying block components with the largest leading eigenvalues.

In asking how common Simon-Ando decomposability may be for generic mutation-selection systems, we work with a Hamming graph configuration space appropriate to the analysis of n-locus, two allele system where only point mutations are allowed in each generation. This representation can be generalized to multi-allelic systems and to models with a small but nonzero probability of multiple mutations per iteration (for example, a mutation matrix with coefficients determined by a Poisson or exponential probability of k mutations in each generation).

Recall that Simon-Ando aggregation and decomposability requires that the vertex set V be completely partitionable into C={$C_1...C_m$} such that every vertex x is an element of some class $C_I$. To be Simon-Ando decomposable, the partition classes are chosen such that intra-partition communication is orders of magnitude stronger than interpartition communication. Applying the Simon-Ando aggregate description (3.13) to the mutation-selection equations (5.1 and 5.5), we have

$$\hat{A}_{IJ} = \sum_{i \in I} \sum_{j \in J} A_{ij} v^*_{jJ} (1_J) = \sum_{i \in I} \sum_{j \in J} M_{ij} w_j v^*_{jJ} (1_J)$$

$$X_I(t+1) = X_I(t) W_I + \sum_{J=1}^{k} M_{IJ} W_J X_J - M_{JI} W_I X_I = \hat{A} X_I(t)$$

where the aggregate variables and parameters are $X_I = \sum_{i \in C_I} x_i$, $W_I = \sum_{i \in C_I} w_i v^*_{iI}(1_I)$, and $M_{IJ} = \sum_{i \in I} \sum_{j \in J} M_{ij}$. Using these parameters, it can be seen that $\sum_I \hat{A}_{IJ} = W_I$.

It follows from the definitions of $\hat{A}$ and the condition for Simon-Ando (3.0) that for all classes J we obtain the following set of inequalities (noting that $W_J \leq 1$, with the assumption that $\epsilon << \frac{1}{2}$)

(6.0)
$$\hat{A}_{JJ} = W_J - \sum_{I \neq J} \hat{A}_{IJ} >> W_J (1 - \epsilon) \implies$$

$$\hat{A}_{JJ} >> \sum_{I \neq J} \hat{A}_{IJ}$$

Rather than making this comparison for every class J, it is instructive to look at the limiting cases, i.e. the set with the highest cross-class communication rate and the one with the lowest within-class communication rate (using the indices 1 and 2 to denote these classes, respectively, and again applying the condition $\epsilon << \frac{1}{2}$):

(6.1 a)
$$\hat{A}_{11} \gg W_1(1-\epsilon) \Longrightarrow$$
$$W_2 - \sum_{I \neq 2} \hat{A}_{I2} \gg W_2(1-\epsilon) \Longrightarrow$$
$$\frac{\hat{A}_{11}}{W_1} > (1-\epsilon) \gg \epsilon > \frac{\sum_{I \neq 2} \hat{A}_{I2}}{W_2}$$

For stochastic matrices, $W_1 = W_2 = 1$ for all J and the same class $C_J$ that minimizes the left hand side maximizes the sum on the right hand side. In the general case, however, the class J which minimizes $\hat{A}_{JJ}$ does not necessarily maximize $\sum_{I \neq J} \hat{A}_{IJ}$, since a high mean partition fitness value leads to relatively high values for both within and cross-partition communication rates. As a stronger condition than (6.1a), we propose

(6.1 b)
$$(\hat{A}_{JJ})_{\min_J} \gg \left( \sum_{I \neq J} \hat{A}_{IJ} \right)_{\max_J}$$

where (6.1b)$\Longrightarrow$(6.1a),(6.0) but not the converse. This inequality can be treated as a limiting case to determine whether a mutation-selection matrix **A** has a Simon-Ando partitioning, as it requires that the lowest-valued within-class communication rate to be greater than the largest-valued cross-class communication rate (the index J on both sides of 6.1b does not generally refer to the same class, the J on the left-hand side refers to the partition class in which within-partition communication is minimal, the J on the right-hand side refers to the partition class with maximal communication with outside classes). This criterion makes heuristic sense, as it follows from (3.0) that all cross-class terms should be of order $\epsilon$, orders of magnitude smaller than the terms which dominate the within-class communication rates.

We propose two different classes of landscape decompositions as candidates for a Simon-

Ando partitioning. The first example is the standard scenario of an adaptive peak and its immediate neighborhood, in the form of a radius $\kappa$, n-dimensional ball (with n the number of loci) centered about a local optimum $x_{0_I}$, to define the class $C_I = \{x | d(x, x_{0_I}) \leq \kappa\}$, with the number of neighbors within and outside the set determined by the topology of a Hamming graph. For the special case of $\kappa = 1$, we have

(6.2 a)

$$\hat{A}_{JJ} = \sum_{i,j \in J} (1 - (n-1)\mu) v_j^* w_j + (n-1) v_0^* w_0 \mu$$

$$= \bar{W}_J (1 - (n-1)\mu) + (n-1) v_0^* w_0 \mu$$

$$\sum_{I \neq J} \hat{A}_{IJ} = \sum_{I \neq J} \sum_{i \in I} \sum_{j \in J} v_j^* w_j \mu = \bar{W}_J (n-1) \mu - (n-1) v_0^* w_0 \mu$$

Similarly for $\kappa = 2$, one needs to take into account "residue" terms for both the Hamming distance 0 and 1 classes with the factor n-2. The sum over the $\kappa = 2$ terms gives a common factor with $W_J$. This suggests a general form for any value of $\kappa < n$

(6.2 b)

$$\hat{A}_{JJ} = \sum_{i,j \in J} (1 - (n-\kappa)\mu) v_j^* w_j M_{ij} + \sum_{\{j | d(x_0, x_j) = \kappa\}} (n-\kappa) v_j^* w_j \mu$$

$$= W_J (1 - (n-\kappa)\mu) + (n-\kappa) \sum_{\{j | d(x_0, x_j) = \kappa\}} v_j^* w_j \mu$$

$$\sum_{I \neq J} \hat{A}_{IJ} = W_J (n-\kappa) \mu - (n-\kappa) \sum_{\{j | d(x_0, x_j) = \kappa\}} v_j^* w_j \mu$$

The inequality (6.1b) compares the minimal valued within-class communication rate with the maximal-valued cross-class communication rate. We will denote these (respectively) as $C_1$ and $C_2$ with corresponding mean fitnesses $W_1$ and $W_2$. Using the values in (6.2b) to determine when (6.1) is satisfied, we get

$$(6.3)$$
$$W_1 (1 - (n - \kappa_1) \mu) + (n - \kappa_1) \sum_{\{j | d(x_0, x_j) = \kappa_1\}} v_j^* w_j \mu >>$$
$$W_2 (n - \kappa_2) \mu - (n - \kappa_2) \sum_{\{j | d(x_0, x_j) = \kappa_2\}} v_j^* w_j \mu$$

The subscripts below the $\kappa$ values indicate the fact that $C_1$ and $C_2$ need not be partitions of the same size, because in general, an n-dimensional hypercube need not be fully partitionable into a set of balls of a fixed radius $\kappa$. Any hypercube where $2^n \neq 0$ mod (n+1) cannot be fully partitioned into radius $\kappa$-balls, and even hypercubes where n+1 does divide $2^n$ cannot always be subdivided into balls of a fixed radius. However, with the assumption that a single vertex is a trivial (radius 0) ball, some partitioning into balls of varying radii is always possible. In most cases, $\kappa_1 \neq \kappa_2$, though there are special cases (for example, if n=3, a partition into $\kappa$=1 balls about 000 and 111) where the lattice can be fully partitioned into balls of the same radius.

Because more terms are added to the left hand side term than those added to the right for larger values of $\kappa$, it is clear that a larger radius about any given point is more likely to give to a Simon-Ando partition. This suggests upper and lower bounds for each side of the inequality, as condition (6.3) will be satisfied for sufficiently small $\mu$ if the following relations hold (obviously, the converse need not be true for sufficiently large values of

$$(n - \kappa) \sum_{\{j | d(x_0, x_j) < \kappa\}} v_j^* w_j \mu )$$

(6.4 a)

$$W_1 (1 - (n - \kappa_1) \mu) \gg W_2 (n - \kappa_2) \mu \implies$$
$$\frac{W_1}{W_2} \gg \frac{(n - \kappa_2) \mu}{1 - (n - \kappa_1) \mu} \implies \frac{W_1}{n(W_1 + W_2) - \kappa_1 W_1 - \kappa_2 W_2} \gg \mu$$

in the special case where $\kappa_1 = \kappa_2$, the inequality is

(6.4 b)

$$W_1 (1 - (n - \kappa) \mu) \gg W_2 (n - \kappa) \mu \implies$$
$$\frac{W_1}{W_2} \gg \frac{(n - \kappa) \mu}{1 - (n - \kappa) \mu} \implies \frac{W_1}{(n - \kappa)(W_1 + W_2)} \gg \mu$$

If we chose a sufficiently small per-site mutation rate $\mu$ (for small values of $\kappa$ and $W_1$ and $W_2$ of roughly the same order, $\mu \ll \frac{1}{n}$), the time scale decoupling characteristic of Simon-And systems will be satisfied for a partition about a local optimum.

In most models of sequence evolution, $\mu$ is chosen to be of the order $\frac{1}{n}$, in which case Simon-Ando aggregation is not likely to give a very close approximation to the system dynamics. In order for the Simon-Ando approximation to hold, the mutation rate must be very low, implying that while there may be a time-decoupling across partitions, the within-class approach to equilibrium will also proceed very slowly. An interesting consequence of this result is that for sufficiently small mutation rates, a "flat" fitness landscape (i.e. one where W(x)=1 for all genotypes x) will be Simon-And decomposable for an arbitrary partitioning (provided that each partition is radius $\kappa$ n-ball about any center vertex). In this special case at least, the time decoupling is due entirely to mutational connectivity within the partitions rather than any fitness differentials within or between partitions.

Our intuition suggested that a fitness landscape structure of several peaks with partition border regions defined by valleys would be necessary to achieve Simon-Ando, in fact this is not the case. A given mutation rate does place constraints on the maximal mean fitness differential across partitions, however, as setting $W_1=1$ and $W_2=1+S$ and rearranging terms in (6.4) implies:

$$S << \frac{1 - 2n\mu - \mu(\kappa_1 + \kappa_2)}{\mu(n - \kappa_2)}$$

again, when $\kappa_1 = \kappa_2$, the relation is

$$S << \frac{1 - 2\mu(n - \kappa)}{\mu(n - \kappa)}$$

Since the parameter is defined so that S>0, this constraint is only meaningful for $\mu$(n-k)<1/2 and $\mu$<<1/n. If these conditions are met, the inequality requires that the mean fitness differential across classes must be less than the term on the right hand side. This places a constraint on the choice of partitions, i.e. one cannot chose one class $C_1$ with a fitness substantially larger than that of $C_2$, at least not for biologically realistic values of $\mu$. Consequently, if a fitness landscape consists of a relatively limited number of high-fitness vertices with the remaining genotypes a low-fitness valley, no partition class can be chosen to consist solely of low fitness genotypes. For example, any partitioning a two-peak landscape with high fitness vertices at (for instance) {00...0} and {11...1} and W(x)<<1 elsewhere must include one of the peaks, limiting the total number of possible partitions to two. The disadvantage of this constraint should be obvious: if the goal of the decomposition and partition approach is to reduce the effective number of state variables, one stands to gain comparatively little in terms of computational efficiency by partitioning a landscape into two blocks.

Furthermore, the above constraints on the value of S require that the differences in fitness betweenthe peaks in different partition classes must not be too great. If there were a large difference in fitness between two peaks, intuition suggests that offspring of the higher fitness region would saturate the entire landscape faster than the lower fitness local peak neighborhoods could attain quasi-equilibrium. This is confirmed by the result that a separation in fitness between partition neighborhoods is inconsistent with Simon-Ando dynamics.

It should be noted that these inequalities place no constraints on the internal structure of each partition class. Intuition might suggest that the "ideal" scenario for Simon-Ando decomposability would be a partitioning where each block consists of a center "peak" surrounded by lower fitness genotypes at the periphery, because we expect lower-fitness genotypes at the edges of a partition to communicate with outside partitions at a lower rate than would high fitness vertices. Though certain within-partition properties may certainly facilitate cross-class separation, as the example of a landscape with all genotypes having a fitness of unity illustrates, there need not be any constraints on within-partition fitness distributions given a sufficiently low mutation rate.

The internal structure of the partition classes may become important for higher mutation rates, i.e. in situations where (6.4) is not satisfied while (6.3), as the more general condition, does hold for a choice of partitions. While (6.4) only places conditions on the mean fitnesses of individual classes, the "residual" terms in (6.3), is

$$(n - \kappa) \sum_{\{j | d(x_0, x_j) < \kappa\}} v_j^* w_j \mu$$

In order for this term to have a significant effect on the relative magnitudes of the two sides of the inequality in (6.3), the product $v_j^* w_j$ must be large for $j | d(x_o, x_j) < \kappa$. For example, in the case of $\kappa = 1$, the fitness and (consequently) the quasiequilibrium frequency of the center optimum must be significantly higher to compensate for the difference. In this situation, the standard "peak and valley" intuition does seem to be consistent with the requirements for decomposability.

An alternative partitioning we investigated in the context of Simon-Ando decomposition is one where each $C_I$ contains members of an equivalence class characterized by common allelic states at any particular locus or subset of loci. One can interpret these equivalence classes as partitionings according to shared "character state" or "schema" identity (sensu Holland 1977, Goldberg 1989, Altenberg 1994). As examples, for a single locus schema, the equivalence classes are $\{0^{**}...^*\},\{1^{**}...^*\}$, for a two locus schema $\{00^{**}..^*\},\{01^{**}...^*\},\{10^{**}...^*\},\{11^{**}...^*\}$, where $^{**}...^*$ represent arbitrary allelic configurations at the remaining loci and the schema loci are chosen to be the first $\kappa$ for the sake of convenience. For a two-allele system, there are $2^\kappa$ equivalence classes (where a decomposition based on the allelic identity at a single locus gives only two classes, while a multilocus schema can be interpreted as giving rise to "nested" sequential partitions over $1,2...\kappa$ site equivalence classes).

It should be obvious that for a $\kappa$-length schema, there are $\kappa$ point mutations that put the offspring into another equivalence class versus n-$\kappa$ possible mutations that remain inside the partition. Applying (6.1a) to these partitions,

$$(6.5) \quad \hat{A}_{JJ} = \sum_{i,j \in J} (1 - \kappa\mu) v_j^* w_j M_{ij} = W_J (1 - \kappa\mu)$$

$$\sum_{I \neq J} \hat{A}_{IJ} = W_J \kappa\mu$$

Note that both the internal and cross class communication rates are in this case independent of the total number of loci, being determined solely by the mutation rate and the number of sites defining a schema. Here we use $C_1$ and $C_2$ to denote the schema classes with the minimal intra-component communication rate and maximal cross-partition communication rate.

In order for Simon-Ando separation to apply, we need

(6.6)

$$W_1 (1 - \kappa\mu) \gg W_2 \kappa\mu \implies$$
$$\frac{W_1}{W_2} \gg \frac{\kappa\mu}{1 - \kappa\mu}, \quad \frac{W_1}{\kappa(W_1 + W_1)} \gg \mu$$

which, along with the expected constraint on mutation rate, implies a stronger separation for smaller $\kappa$ (fewer loci defining fewer partitions). Rewriting (6.6) in terms of selection parameter S, where $W_1 = 1-S$, $W_2 = 1$, the inequality is equivalent to $S \ll \frac{1}{1-\kappa\mu}$. For $\kappa\mu \sim 1/n$, this is $S \ll \frac{n}{n-1}$ (implying that for large numbers of loci, no more than a twofold difference in mean partition fitness is permissible).

The mean fitness $W_J$ can be interpreted as the marginal fitness of the particular schema configuration defining the Jth partition class. For example, in a twofold landscape partition defined by 1**...* versus 0**...*, $W_J$ represents the mean (marginal) fitness of allele 1 or 0 at the first locus. Because aggregability of the partition implies the (near) dynamical sufficiency of mean fitness and effective transition rate parameters, the condition (6.6) can be seen as indicating when schema can be treated as "units of selection" (see below). For this condition to be satisfied, the internal cohesion within an equivalence class must be greater than the communication from outside partitions, and as the inequality above indicates, this holds only when the mean (marginal) fitness of any particular schema configuration is of approximately the same order as the mean fitness of the other configurations.

A series of numerical experiments (computed using *Mathematica*, with programs available from the senior author on request) confirm that the landscapes induced by mutation-selection matrices satisfying (6.1b) are aggregable and decomposable. What is somewhat open-ended given the constraints on mutation rates and fitness differences in (6.4b) etc. is just how small the per-locus mutation rate $\mu$ has to be with respect to the ratio on the left-hand-side. It turns out that for certain landscapes, the Simon-Ando approximation to $x_{i_I}(t)$ and $v_{i_I}^* X_I(t)$ (and therefore, approximations of stationary distributions $v_{i_I} = v_{i_I}^* \tilde{X}_I$, with $\tilde{X}$ the stationary distribution for $\hat{A}$) breaks down for all but very small mutation rates, while for others the results are robust for biologically realistic transition rates.

We begin with the special case of a system where the fitnesses of every genotype are identically set to unity. Using a five locus, two allele system as an example, we (for convenience) chose the equivalence classes to be all of the vertices within Hamming distance 2 from {00000} and {11111} respectively. Clearly, the within-class quasi-equilibria $v_I^*$ will approximate uniform distributions irrespective of mutation rates, with the global distribution over the $T_1 < t$ time interval approximated by the product of initial within-partition aggregate frequencies and the within-class uniform distributions. Choosing the mutation rate to be $\mu=0.01<<0.25$, we compare the equilibrium genotype distributions $v_{i_I}$ (first eigenvector) of the mutation-selection matrix **A** to the aggregation of variables estimates $v_{i_I}^* \tilde{X}_I$ in Figure 1. The mean-square error is approximately 0.08, which is a reasonable approximation (and one which doesn't improve with lower mutation rates, reflecting the fact that 6.4b gives a sufficient but not necessary condition for time-separation).

In the next set of figures, we repeat the same calculations, but in this case we have multiplicative fitness functions about each local optimum, so that the fitness of any genotype is $W(x) = W_0 (1-s)^{d_{\min}(x, x_0)}$ with $x_0$ the nearest local optimum and s=0.1. The mutation rate is set to $\mu=0.01$, while the peak fitnesses are chosen to be {1,1} in Figure 2 and {1,0.9} in Figure 3. The first case corresponds to the "degenerate quasispecies" discussed by Eigen and Schuster (1989) and Schuster and Swetina (1988), with identical eigenvalues $\lambda_1 = \lambda_2 = .990181$ where the associated eigenvectors have most of their probability density about either peak (Fig 2a). The stationary distribution (plotted in Figure 2b) can be constructed as a weighted superposition of eigenvectors associated with the leading degenerate eigenvalue.

Aggregation of variables gives a very good approximation to the stationary distribution for this fitness landscape. The values of $v_{i_I}* \tilde{X}_I$ are plotted against the stationary distribution in 2b. The mean-square error of $v$ versus $v_I* \tilde{X}_I$ is of the order of $10^{-2}$, a value that improves for lower values of $\mu$ (e.g. $\mu$=0.001 gives errors of the order of $10^{-5}$. This is consistent with our expectation, since the Simon-Ando approximation improves for lower transition rates.

The results are similar for the landscape where the two peaks differ in height, apart from the fact that the eigenspace is non-degenerate. The normalized leading eigenvector (corresponding to the equilibrium distribution) is plotted against $v_I* \tilde{X}_I$ in Figure 3, again giving a close match with a mean-square error at the order of $10^{-2}$. We note that the approximations are good whether we use the representation of aggregate transition rates (1.1) or (3.13).

The situation becomes more interesting in the case of two-peaked landscapes with nearly identical peak fitnesses. Consider a scenario where one peak has a fitness of unity while the other has a value close to unity, i.e. 0.999. It was shown by Schuster and Swetina (1988) and Wilke (2001b) that for sufficiently high mutation rates the equilibrium distribution depends on the fitnesses of the immediate mutational neighbors of each peak. Namely, if the more fit peak is surrounded by relatively low fitness neighbors while the peak with slightly lower fitness has comparatively high fitness neighbors (corresponding to a "mutationally robust" or "genetically canalized" genotype, e.g. Wagner et al 1996), the equilibrium density will often be concentrated at the somewhat lower "plateau" rather than the sharp, isolated peak.

We replicated these results in Figure 4, which depicts the equilibrium genotype distribution for a fitness function defined by $W(x) = W_1 (1 - s_1)^{d_{\min}(x, x_1)}$ about the first peak and

$W(x) = W_2(1-s_2)^{d\min(x,x_2)}$ about the second peak, where $s_1=0.9$ and $s_2=0.1$ and $x_1$, $x_2$ are the local peak vertices (representing a fitness landscape with one sharp peak and a relatively flat fitness plateau). The fitness values are chosen such that the peak is slightly higher than the plateau, i.e. $W_1=1.0$, $W_1=.99$.

The first computations are for a relatively high mutation rate of $\mu=0.1$, which as expected gives a stationary distribution at the plateau. If we approximate the aggregate dynamics as (1.1), the estimate of the leading eigenvector is completely misleading, specifically, it predicts a stationary distribution concentrated at the peak. However, using (3.13) to estimate the leading eigenvector does predict a distribution concentrated at the "plateau," although it can be readily seen in Fig 4 that the match is not very close (the mean square error in this case is nearly 0.2). This is due to the fact that if we use $\mu=0.01$, the stationary distribution is still concentrated at the plateau and the aggregate distribution is somewhat improved, being just under 0.1.

For substantially smaller (and biologically unrealistic) mutation rates such as $\mu=0.001$, the stationary distribution is always concentrated at the highest peak (because, as Schuster and Swetina demonstrated, selection pressure in favor of the lower plateau is a second order effect that scales in proportion to mutation rate, when mutation rates are sufficiently low, peak genotype offspring rarely "encounter" their neighbors). For this case, Simon-Ando aggregation gives very good estimates (Figure 4b), giving square error values of the order of $10^{-3}$. The estimates are good both for (1.1) and (3.13) as operators for the aggregate dynamics.

The applicability of Simon-Ando decomposability to these model landscapes has some interesting implications for the way one thinks about units of selection. It was suggested in Shpak et al (2003) that aggregation of variables offers a natural approach to identifying units of selection above the genotype level, in that aggregation methods involve partitioning the genotype space into coherently interacting subsets. Indeed, one criterion proposed for identifying units of selection is that of dynamical sufficiency (see Lewontin 1970, Wimsatt 1981).

While a case can be made that landscape partition classes behave as coherent entities, it doesn't seem to be the case that they must necessarily do so as a consequence of Simon-Ando dynamics. For example, Phase III of Wright's "shifting balance" model (1932) in which interdemic selection (aggregate competition between sets of genotypes composing more or less isolated peaks and their neighborhoods) is actually incompatible with a Simon-Ando model, because mean fitness differences between neighborhoods are in violation of (6.4b). Yet interdemic selection (as a special case of group selection) has been shown to be at least in principle a possible, if not necessarily an important, evolutionary force (Coyne et al, 1997).

Similarly, the example of a landscape with an isolated high fitness peak competing against a lower fitness but more "robust" genotypes is a case in point - the parameter values where the plateau dominates the distribution are also those where the Simon-Ando approximation starts to break down due to relatively high mutation rates. Yet the long-term behavior of such a system can be predicted from the fact that the eigenvalues associated with the plateau are larger than those associated with the isolated peak under the right choice of parameters (Schuster and Swetina 1988, Wilke 2001), which suggests that in some sense of the word the competition is taking place between the peak and plateau as aggregate entities. This suggests that one can have higher-order entities that act coherently under selection whose aggregability is not necessarily a result of the fast-slow dynamics. As a case in point, competition between two-peaked landscapes was analyzed in the context of unweighted equitable partitions whose aggregativity was exact and not time dependent, contra Simon-Ando.

We now turn our attention to another issue in the units of selection question that can be elucidated by Simon-Ando aggregation and decomposition: the problem of character identification and schema-based identity classes.

## ■ Schema Partitioning and the Definition of Evolutionary Characters

In our earlier paper (Shpak et al 2003), we compared the formal property of landscape decomposability into equitable partitions defined by fitness and mutational distance equivalence classes to the character state decomposition models developed in Wagner and Laubichler (2000a,b). The latter form of decomposability defines equivalence classes in terms of identical allelic states at a locus or over subsets of loci (schema). It can be seen that in general equitable partitions need not correspond to schema equivalence classes.

However, there are cases where an equitable partitioning does correspond to a partitioning into schema equivalence classes. In the previous section we examined a partitioning defined by schema which for the appropriate choice of mutation rate and fitness function (6.6) satisfies the condition for Simon-Ando decomposability (which, in turn, is a special case of weighted equitable partitioning). Each schema equivalence class and its frequency can be interpreted as a "character state," and when the mutation-selection system is nearly decomposable, the frequencies of the equivalence classes are a dynamically sufficient descriptor of the evolutionary process. Since one condition for identifying units of evolution is determining whether the entities in question are dynamically sufficient, it is natural to equate selection on a schema class with selection for an independent character state.

What then is the relationship between Simon-Ando decomposability of schema partitions and the character partitionings of Wagner and Laubichler? To recapitulate Wagner and

Laubichler's results (with changes made to the notation for consistency with this paper, as well as some formal adjustments for a discrete-time representation), define a set of genotypes $\{x_1....x_N\}$ with an associated frequency vector $p_i=\{p_1...p_N\}$. The equivalence classes $C^1=\{C_1....C_k\}$ and their associated frequencies $\pi_I=\{\pi_1...\pi_m\}$ are defined such that every genotype $x \epsilon C_I$ has an identical allelic state at a particular locus. More generally, the equivalence classes are defined as a set of genotypes identical over some arbitrary subset of sites, or a "schema" (sensu Holland 1975, Goldberg 1988, Altenberg 1995). One such equivalence class (defined as an allele at a single locus) for a 4-locus genotype would be the set of all genotypes $C_1$ of the form 0***, $C_2$ of those of the form 1***, defining partition C. In turn, another class of partitions $C^2$ will be defined by the allelic identity at the second locus, and so forth.

Wagner and Laubichler define the Cartesian product $C^1 \times C^2$ to be an oc (orthogonal compliment) partitioning if $C=C^1 \times C^2$, or more generally, $C=C^1 \times C^2...\times C^v$. For example, $C=C^1 \times C^2$ could be the Cartesian product of allelic variants at loci 1 and 2, respectively. The authors construct an oc-partition by choosing a set of invertible functions $F=\{f_{ij}|f:C_i \rightarrow C_j\}$ which maps every element in one equivalence class to the corresponding genotype in another class. For example, f could map 101 to 001, with *01 defining an (orthogonal) equivalence class with respect to the first locus. In the case where F is a family of transitive maps, i.e. $s=f_{IJ}(x)$ and $t=f_{JL}(u)$ implies $t=f_{IL}(x)$, F defines a complementary (orthogonal) partitioning $\overline{C}=\{\overline{C_1}...\overline{C_k}\}$, where every class is in the complementary partition is $\overline{C}=\{s \simeq x$ if there is $f_{IJ} \epsilon F|s=f_{IJ}(x)\}$. This mapping specifies an equivalence relation because the functions in f are transitive and invertible (Rosen 1984, Bogart 1990), and consequently defines each genotype x as $x=C_I \cap \overline{C_J}$.

Given an oc-partitioning, Wagner and Laubichler have shown that for fitness functions satisfying the additivity condition and for "character" frequencies satisfying a generalized linkage equilibrium, the equivalence class frequencies $\pi_I$ are a dynamically sufficient descriptor of evolution under a selection operator.

Carter (1997, unpublished) has shown that the equivalent condition in discrete time requires that all fitness functions are multiplicative across partitions, with w the Malthusian fitness parameter:

(6.7)  $w(f_{IJ}(x)) = c_{IJ} w(x)$

In other words, the fitness differences between members of the same equivalence class (i.e. allelic state at a particular locus) are determined by a constant product term $c_{IJ}$ determined by the rest of the genotype or character state configuration. This effectively excludes any type of nonlinearity due to epistasis in fitness functions. The other condition, of course, is generalized linkage equilibrium,

(6.8)  $p_I(f_{IJ}(x)) = p_J(x)$, where $p_J(x) = \dfrac{p(x)}{\pi_J}$

with p(x) denoting the frequency of x while $p_J(x)$ refers to the marginal frequency in the Jth partition. This definition is equivalent to the conventional linkage equilibrium condition $p(x \in \pi_I \cap \pi_J) = \pi_I \pi_J$.

It was shown in Carter's derivation that if these criteria are met, the selection equation on genotypes (here in discrete time)

(6.9 a)   $p_i(t+1) = p_i(t) w_i$

can be aggregated into a dynamically sufficient description as

(6.9 b)   $\pi_I(t+1) = \pi_I(t) \bar{w}_I$

The condition required for these to hold is that the within-class variances in fitness $V_I$ be the same across partition classes $C_I = C_1 \ldots C_m$. This condition turns out to be generally stable only under multiplicative fitness and generalized linkage equilibrium.

In order to make a meaningful comparison between character partitioning under generalized linkage equilibrium and Simon-Ando, it is necessary to determine under which models of transmission (mutation) the within-class variances (and by extension, the generalized linkage equilibrium) are remain stable. In an unpublished manuscript, Altenberg (2002) proposed that for two orthogonal partitions $C^1$ and $C^2$ the following tensor product definition of generalized linkage equilibrium is stable under multiplicative fitness functions and mutation operators:

$$(6.10) \quad x(t) = x^1 \otimes x^2 = \left(x_i^1 \, x_j^2\right)_{i=1\ldots N_1,\, j=1\ldots N_2}$$

where $\otimes$ is the Kronecker product operator $(A \otimes B)_{ij} = AB$. In vector form, the within-partition frequency distributions x are given by $x^1 = (I_1 \otimes 1_2^T)x$ and $x^2 = (1_1^T \otimes I_2)x$, where **I** is an identity matrix and **1** is a vector of 1's corresponding to the number of orthogonal classes in the respective partitionings.

Rewriting the transition operator **A**=**WM**, with the diagonal fitness matrix **W** and **M** the mutation matrix, in order to satisfy the multiplicativity condition,

$$(6.11) \quad W = W_1 \otimes W_2$$

and a similar structure for the mutation operator across two partitionings

$$(6.12) \quad M = M_1 \otimes M_2$$

which can be extended to an arbitrary number of orthogonal partitions as $W = W_1 \otimes W_2 \ldots \otimes W_m$ etc. An interpretation of (5.6) is that mutation rates are independent at each locus (or partition) irrespective of the identity at the other loci.

To show that a Kronecker product model of mutation and fitness effects conserves partition independence and generalized equilibrium, note that

$$x(t+1) = (W \otimes W)(M \otimes M)\left(x^1(t) \otimes x^2(t)\right) =$$
$$(MW)\, x^1(t) \otimes (MW)\, x^2(t) = x^1(t+1) \otimes x^2(t+1)$$

implying that x,x are dynamically sufficient descriptors, since (5.7) implies that each equivalence class frequency trajectory can be represented as

$$x^1(t+1) = W_1 M_1 x^1(t),\quad x^2(t+1) = W_2 M_2 x^2(t)$$

This proves that generalized linkage equilibrium $x = x^1 \otimes x^2$ is stable given multiplicative fitness and mutation rate. The results can be summarized as saying that dynamical sufficiency of schema frequencies (i.e. "character decomposability") requires independent mutation rates and multiplicative fitness across partitions as well as multiplicativity of cross-partition frequencies.

This is in contrast to the conditions under which a schema-based partitioning is decomposable according to the Simon-Ando criterion. Simon-Ando decomposition places no conditions on the frequency distribution apart from the requirement that the distribution within a partition class be in quasi-equilibrium (which, in general, would generate strong linkage disequilibria). Consequently, we can conclude that the two cases in which schema frequencies are dynamically sufficient as evolutionary "characters" are entirely independent of one another, suggesting at least two different sets of conditions (i.e. 6.6 versus 6.11-12 combined with linkage equilibrium) where schema can modelled as units of evolution.

If aggregativity and decomposability into classes defined by Hamming distance suggests units of selection above the genotype level, there are interesting conceptual implications for inter-preting a partitioning based on schema equivalence classes. Traditionally, in any system where schema (e.g. allelic) frequencies are dynamically sufficient, evolution is thought of as occurring at a level below that of the genotype. In particular, when allele frequencies are a dynamically sufficient descriptor, it is often stated that the "gene" is the unit of evolution.

In reality, in both cases where schema frequencies are dynamically sufficient (i.e. the Wagner-Laubichler criterion and Simon-Ando), the aggregation and decomposition procedures are formally equivalent to the $\kappa$-ball partitions in that they involve treating sets of genotypes as aggregate entities. In other words, if we are to consider the case where $\kappa$-ball partitions are dynamically sufficient as selection on a set of genotypes ("group selection" in some broad sense), then it is not meaningful to think of the case of dynamically sufficient schema classes as selection below the level of the genotype. Quite the contrary, allelic or schema frequencies are simply a book-keeping shorthand for the frequencies of an aggregate class, and it is the symmetries or fitness equivalences within these classes that make the "gene frequency" shorthand possible. Far from confirming that selection occurs at the genic level, the dynamical sufficiency of schema frequencies actually lends support to the idea of units of evolution above the genotype level.

## ■ Recombination and Simon-Ando Decomposability

The constraints on which mutation-selection systems have Simon-Ando dynamics are a general result of the neighborhood relations (specifically, the Hamming graph) induced by point mutation acting on Boolean or $\alpha$-alphabet strings. It is natural to ask whether Simon-Ando dynamics can arise under different transmission operators, namely recombination. It has been shown elsewhere (Gitchoff and Wagner 1997) that the configuration space $\chi$ induced by recombination is a hypergraph with "edges" consisting of intermediate recombinant classes. Here we ask whether Simon-Ando decomposability can exist given a recombination-induced configuration space. It seems intuitive that Simon-Ando decomposability may be more difficult to achieve because a much wider spectrum of possible offspring genotypes can be produced under recombination than under point mutation, thus reducing the internal cohesion of any neighborhood class of genotypes.

Both the neighborhood relationships and the transition probabilities attached to these edges differ from any model of mutation in that they depend on which partner genotypes are present in the population. As a result, the transition rates $A_{ij}$ are generally dependent on the probability of encountering of recombinant partner k in the population. Under random mating the transition rate from j to k is given by $A_{ij} = \sum_{k=1}^{n} x_k T_{jk \to i}$, where $T_{jk \to i}$ is the probability that a pairing of genotypes j,k produces recombinant offspring i. Under these assumptions the recombination-selection equations are

$$(7.1) \quad x_i(t+1) = \sum_{j=1}^{N} \sum_{k=1}^{N} w_j w_k x_j x_k T_{jk \to i}$$

Again, we use absolute frequencies and Malthusian fitnesses to avoid the complications of a mean fitness term in the denominator. If mating is nonrandom, the most general representation replaces $x_i x_j$ above with $\phi(x_i, x_j)$, which is the joint probability of the i,j pairing.

Under the action of a point mutation operator, the neighborhood sets of any genotype were definable as the point mutational neighbors (i.e. all i neighbors of j if $\mu_{ij} \neq 0$, or $\mu_{ij} < \epsilon << 1$), a similar definition of recombination neighbors based on $\sum_{k=1}^{n} x_k T_{jk \to i}$ is less straightforward because it depends on the distribution of parental genotypes in the population. In general, a crossover operator involves a function on vertex (genotype) pairs $\chi : (V,V) \to (V,V)$, in contrast to the mutation operator $\xi : V \to V$.

The general structure of a recombination operator can be described as follows: given a length n string with a $\alpha$-letter alphabet ($\alpha=2$ for Boolean strings), we represent the crossover operator $\chi$ by denoting the kth site of genotype y as $y_k$, which gives us $\chi(y,z)=(u,v)$, $\forall k:(y_k=u_k \wedge z_k=v_k) \vee (z_k=u_k \wedge y_k=v_k)$, i.e. at all sites k, the recombinant takes on the site identity of parent y while the sister (complementary) recombinant takes on the site identity of parent z. An arbitrary recombination operator R can be represented as a mapping from the set of vertex pairs onto the power set of V, i.e. $R:(V,V) \rightarrow P(V)$, hence the P-structure terminology in Stadler and Wagner (1999).

For an operator $R_1$ associated with a single point recombination event, recombination at the kth point gives us (following the notation of Stadler et al 2000, Stadler and Wagner 1999):

$$R_1(y,z) = \chi_k(y,z) = \{(y_1 \ldots y_k z_{k+1} \ldots z_n), (z_1 \ldots z_k y_{k+1} \ldots y_n)\}$$

This definition of the offspring set can obviously be generalized to arbitrarily many crossover points. The limiting case given a high per-site recombination rate is one where any number of recombination points from none to all n loci are equally possible (i.e. "uniform crossover"), whereby each site k in a given offspring has an equal likelihood of coming from either parental genotype. The action of the uniform recombination operator $R_U$ on two parental strings is described by:

$$R_U(y,z) = \{(v_1 \ldots v_n): v_i = y_i \text{ or } z_i \ \forall i\}$$

Given a Hamming distance $H(y,z)=d$ between the two parental sequences, it can be shown that the size of the recombination sets is (Gitchoff and Wagner, 1996):

$\|R_1(y,z)\| = 2d$ when $y \neq z$, 1 if $y=z$

$\|R_U(y,z)\| = 2^d$

For the sake of computational simplicity, we will only examine the cases of uniform and single-point recombination, as they are the limiting cases under very high and relatively low cross-over rates.

We now turn to the question of whether recombination systems can satisfy the conditions for Simon-Ando near-decomposability, i.e. whether there exist fitness landscapes that induce a fast-slow dynamic and local quasiequilibria under the action of recombination. The existence of such quasi-equilbria implies that in principle an aggregation of state variables and parameters is possible, allowing the construction of dynamically sufficient representation of aggregate recombination dynamics

$$X_I(t+1) = \sum_J \sum_K W_J W_K X_J X_K T_{JK \to I}$$

where we define $X_I = \sum_{i \in I} v^*_{i_I}$ and $W_J$ as the mean fitness within the partition $\sum_{i \in I} w_i v^*_{i_I}$, the "aggregate" recombination rate across partitions is estimated by

$T_{JK \to I} = \sum_{i \in I} \sum_{j \in J} \sum_{k \in K} v^*_{i_I} v^*_{i_I} T_{jk \to i}$.

It should be noted that since (7.1) is a quadratic dynamical system (e.g. Rabinovich et al 1992, Rabani et al 1995), there is no reason to expect that either the inequalities defining Simon-Ando type properties or the aggregation rules for linear systems should apply here. However, since one of the defining properties of a Simon-Ando dynamic is the existence of local quasi-equilibrium distributions, if there exists such as distribution $p^*$ for evolution under recombination and selection, then we can approximate $p(t+1) = f(p(t))$ as a linearization about $p^*$.

It is important to remember that any statements about the structure of the linearized transition operator **A'** do not allow one to make a statement about whether the recombination-selection system will have Simon-Ando type fast-slow dynamics. In order to do so, one needs information about the long-term behavior of the system, one in which the distribution of state variables can be quite far removed from any particular choice of distribution for the local linearization. So in stating that the linearized representation is "Simon-Ando" or that it satisfies (6.1b), the most that can really be said is that the local behavior of the system at some particular distribution is consistent with Simon-Ando in the short term (for instance, one can infer the near-stability of local quasi-equilibria).

However, we propose that the local linearization serves as a heuristically useful tool for identfying systems that are not Simon-Ando as well as those which satisfy the local quasi-equilibrium properties associated with the first phase of Simon-Ando dynamics. If even the local linearization fails to have Simon-Ando structure, then clearly the system will not have Simon-Ando type dynamics globally. Conversely, if the local linearization gives a transition operator consistent with Simon-Ando dynamics, then one can argue that at least the first stage of Simon-Ando decomposability, i.e. the existence of local quasi-equilibria, may be fulfilled.

Linearizing about the proposed quasi-equilibrium distribution, we obtain,

(7.2)
$$p_i(t+1) = f_i(p(t)) = \sum_{j=1}^{N} p_j w_j \sum_{k=1}^{N} p_k w_k T_{jk \to i}$$

$$p_i(t+1) \approx p_i^* + \sum_{j=1}^{N} \frac{\partial f_i}{\partial p_j}(p_j^*)(p_j - p_j^*) + O(2) + \ldots$$

$$= p_i^* + \left( \sum_{j=1}^{N} 2 w_j \sum_{k=1}^{N} p^*_k w_k T_{jk \to i} \right)(p_j - p_j^*)$$

in the transformed coordinate system with the perturbation state variables $u=(p(t)-p^*)$, the state dynamics of $u(t)$ are described in terms of the linear operator **A'** where

$$A'_{ij} = 2 w_j \sum_{k=1}^{N} p^*_k w_k T_{jk \to i}$$

can be evaluated given fitness and frequency distributions (as well as models of recombination) to determined whether the system locally induces fast-slow dynamical behavior.

Applying the corresponding sums on both sides of corresponding to inequality (6.1b) to **A'**, we ask under which fitness functions and recombination rules the following will hold:

(7.3)
$$\left( \sum_{i \in J} \sum_{j \in J} 2 p^*_j w_j \sum_{k=1}^{N} p^*_k w_k T_{jk \to i} \right)_{\min J} \gg$$

$$\left( \sum_{I \neq J} \sum_{i \in I} \sum_{j \in J} 2 p^*_j w_j \sum_{k=1}^{N} p^*_k w_k T_{jk \to i} \right)_{\max J}$$

for some distribution $p^*$ (where min/max J refer to the classes J which respectively minimize and maximize the corresponding sides of the inequality). One can chose a "best case scenario" for this condition to be satisfied, namely, one where the fitness values for partition corresponding to the maximal value equal to those corresponding to the minimal value, without any assumptions about the structure of distribution $p^*$ as a limiting case.

As for point mutation, we investigate the decomposability properties with respect to two types of partitions; first the partitioning defined in terms of schema equivalence classes, second the radius $\kappa$ n-dimensional balls around a fixed vertex point. For partitions defined by schema, the case of uniform (free) recombination is relatively straightforward to analyze, because the transition probabilities across schema depend only on the partial Hamming distances across the subset of loci defining the schema.

Specifically, if we rewrite (7.3) in the expanded form:

$$\sum_{i \in J} \sum_{j \in J} 2 p^*_j w_j \sum_{k=1}^{N} p^*_k w_k T_{jk \to i} = \sum_{j \in J} 2 p^*_j w_j \sum_{k=1}^{N} p^*_k w_k T_{jk \to i \in J}$$

and

$$\sum_{I \neq J} \sum_{i \in I} \sum_{j \in J} 2 p^*_j w_j \sum_{k=1}^{N} p^*_k w_k T_{jk \to i} =$$

$$\sum_{j \in J} 2 p^*_j w_j \sum_{K} \sum_{k \in K} p^*_k w_k T_{jk \to i \notin J} =$$

$$\sum_{I \neq J} \sum_{j \in J} 2 p^*_j w_j \sum_{K=0}^{n_S} \sum_{k \in K} p^*_k w_k T_{jk \to i \in I}$$

where for an allelic alphabet of size 2 there are $2^{n_S}$ partition classes $C_I$ given $n_S$ loci defining a schema, each class containing $2^{n-n_S}$ genotypes. Through abuse of notation, the indices K correspond to partial Hamming distance classes where all members of the Kth partition are of partial Hamming distance K to the schema loci in the Jth class (with K=0 corresponding to the Jth class itself). Under free recombination, for any two recombinant genotypes j,k with partial Hamming distances $K=d_S$ the probabilities $T_{jk \to i \in J} = \left(\frac{1}{2}\right)^{d_S}$, while This is simply the probability that in any pairing of distance $d_S$ over the schema loci all of the parental types in schema class J are transmitted to the offspring. Therefore, we have

$$\hat{A}_{JJ} = 2 \sum_{j \in J} p^*_j w_j \sum_{K=0}^{n_S} \sum_{k \in K} p^*_k w_k \left(\frac{1}{2}\right)^K =$$

$$2 W_J \sum_{K=0}^{n_S} \left(\frac{1}{2}\right)^K W_K = 2 W_J^2 + W_J \sum_{K=1}^{n_S} \left(\frac{1}{2}\right)^K W_K$$

$$\sum_{I \neq J} \hat{A}_{IJ} = 2 \sum_{j \in J} p^*_j w_j \sum_{K=0}^{n_S} \sum_{k \in K} p^*_k w_k \left(1 - \left(\frac{1}{2}\right)^K\right) = 2 W_J \sum_{K=0}^{n_S} \left(1 - \left(\frac{1}{2}\right)^K\right) W_I$$

it can be seen that for any choice of $W_1$ and $W_2$ (defined as before as the mean fitnesses of partitions $C_1$ and $C_2$, the subsets with the minimal within-partition and maximal cross-partition communication rates, respectively), the Courtois inequality will not be satisfied under free recombination, because the magnitude of $\sum_{I \neq 2} \hat{A}_{I2}$ will be of the same order as that of $\hat{A}_{11}$. If on the other hand we posit a situation where each genotype has a certain probability $\rho$ of undergoing recombination in every generation versus probability 1-$\rho$ of reproducing through selfing, with overall genotype transition rates

$$p_i(t+1) = f_i(p(t)) = (1-\rho)p + \rho \left( \sum_{j=1}^{N} p_j w_j \sum_{k=1}^{N} p_k w_k T_{jk \to i} \right)$$

which, linearized about p gives the local transition operator A' as

$$A'_{ij} = \rho \left( 2 w_j \sum_{k=1}^{N} p^*_k w_k T_{jk \to i} \right) \qquad i \neq j$$

$$(1-\rho) + \rho \left( 2 w_j \sum_{k=1}^{N} p^*_k w_k T_{jk \to i} \right) \qquad i = j$$

From the above, the within and cross class transition rates are approximated by:

$$\hat{A}_{JJ} = W_J \left[ (1-\rho) + 2\rho \sum_{K=0}^{n_S} \left(\frac{1}{2}\right)^K W_K \right]$$

$$\sum_{I \neq J} \hat{A}_{IJ} = 2\rho W_J \sum_{K=0}^{n_S} \left(1 - \left(\frac{1}{2}\right)^K\right) W_K$$

Analogous to mutation-selection systems with very low mutation rates, it is relatively easy to satisfy (6.1b) given a sufficiently small value of $\rho$, except in cases where $W_2 \gg W_1$. This is to be expected because of the way the model is formulated; as with mutation, offspring in each generation tend to remain in the same partition as parents simply because for very low $\rho$ each genotype produces offspring identical to itself. This form of internal stability is the trivial limiting case for which strict sense "Simon-Ando" decomposability can apply even on a flat fitness landscape (though the degree of inequality is certainly enhanced by the right choice of mean fitness differentials).

For single point recombination the combinatorics of the problem become much more problematic. While under free recombination only the partial Hamming distances between recombinants are needed to calculate the probability of remaining in the parental partition, in the case of single (or 2...n-1 point) recombination requires information about which particular loci determine the partial Hamming distances and their lengths apart on the chromosome. For example, under single point recombination the probability that a recombinant between {11000,00000} produces an offspring outside the 11*** equivalence class is less than the same for a schema defined by 1****1, i.e. {10001,00000} because there are more possible recombination events that could produce offspring outside the 1****1 class than for 11***.

If the equivalence classes are defined by a single locus, calculating the within and cross-partition transition rates $T_{jk \to i \in J} = \frac{1}{2n}$ (for j∈J and k∉J), as the probability of recombination occurring at that particular locus is simply the reciprocal of the number of loci, and given that a recombination even occurs there one half of the progeny will be of the parental type. For a schema defined by two loci, $T_{jk \to i \in J} = \frac{1}{2n}$ for a partial Hamming distance of unity, while for a partial Hamming distance of two, the probability of obtaining a parental recombinant is $\frac{n-D}{2n}$, where D is the number of loci separating the schema elements. In the case where $d_S(j, k) = 0$, it is obvious that $T_{jk \to i \in J} = 1$.

From this we can calculate both sides of the Courtois inequality for the two-locus schema under single point recombination. The relevant equivalence classes with respect to a particular schema class have the partial Hamming distances 0,1,2 based on how many schema loci have different allelic states than the parent genotype (the same procedure with one fewer partial Hamming class can be used to calculate the terms in the trivial case of a single locus schema).

Below $W_2$ and $W_2$ are the mean fitnesses of the partial Hamming distance 1,2 classes with respect to partition J:

$$\hat{A}_{JJ} = 2 \sum_{j \in J} p^*_j w_j \sum_{K=0}^{2} \sum_{k \in K} p^*_k w_k T_{jk \to i \in J} = 2 W_J \left[ W_J + \frac{1}{2n} W_1 + \frac{n-D}{2n} W_2 \right]$$

$$\sum_{I \neq J} \hat{A}_{IJ} = 2 W_J \left[ \left(1 - \frac{1}{2n}\right) W_1 + \frac{n+D}{2n} W_2 \right]$$

The Courtois inequality is not readily satisfied in this case. While $\sum_{I \neq J} \hat{A}_{IJ}$ can be minimized by assuming $W_J \gg W_0, W_1$, the inequality requires a comparison of $\hat{A}_{JJ}$ minimal over all J to $\sum_{I \neq J} \hat{A}_{IJ}$ maximal over all J. Thus minimizing W,W with respect to the maximum value for $\sum_{I \neq J} \hat{A}_{IJ}$ also reduces the magnitude of $\hat{A}_{JJ}$, because the partition J which minimizes the latter is by definition in the 0,1 partial Hamming classes with respect to $J_{max}$. From this we can conclude that (local) Simon-Ando is not likely to occur under single point recombination in the absence of some model which restricts the frequency of recombination itself.

It should be apparent that for longer schema, calculating these transition probabilities will require information about which schema loci constitute the partial distances as well as pairwise (chromosome length) distances for each of those loci. As such, we have no closed-form solution for within and cross-partition communication rates under single-point recombination, though these calculations should be fairly straightforward numerically for sufficiently short schema.

Turning now to partitioning defined by Hamming distance $\kappa$ radii with respect to some choice of reference vertices (the first type of partitioning investigated in the context of mutation-selection systems), chose partition $C_J$ as the set of genotypes within Hamming distance $\kappa$ of $x_0$. To calculate $\sum_{i \in J} A_{ij}$, the probability that the progeny i of j$\epsilon$J are elements of the same partition, one needs the following set of Hamming distances: $d_{0\,j}$ (distance between reference vertex and parent j), $d_{0\,k}$ (distance between reference vertex and parent k), and $d_{kj}$.

From these quantities one can calculate the "shared" distance from the reference vertex to both parents: $k_{jk}=(d_{0\,k}+d_{0\,j}-d_{jk})/2$ (i.e. if the reference vertex consists of all 0's, $k_{jk}$ denotes the number of shared 1's in both parent sequences). In turn, the Hamming distance from the reference genotype to the offspring is $d_{0\,i}=k_{jk}+m_{jk}$, with $m_{jk}$ being the number of offspring loci with allelic states different from the reference vertex which have been inherited from either parent j or k (but not both). This quantity has a binomial distribution given free recombination,

$$P(m_{jk}) = \binom{d_{jk}}{m_{jk}} \left(\frac{1}{2}\right)^{d_{jk}}$$

from which one can derive the probability that offspring i will be within Hamming distance $\kappa$ of the reference vertex (and thus within the partition),

$$T_{jk \to i \in J} = P(d_{0\,i} \leq \kappa) = \sum_{k=0}^{\kappa - k_{jk}} \binom{d_{jk}}{k} \left(\frac{1}{2}\right)^{d_{jk}}$$

Unfortunately, there is no actual closed form solution for this partial sum, which evaluates to the following, where $H_{21}$ is the Kummer confluent Hypergeometric function of the second kind, i.e. $H_{21}(a,b,c,z)=\sum_{k=0}^{\infty}(a)_k (b)_k /(c)_k \, z^k/k!$,

$$P(d_{0\,i} \leq \kappa) = 1 - \frac{d_{jk}\left(2^{-d_{jk}}\right) d_{jk}! \, H_{21}\left(1, 1+\kappa-k_{jk}-d_{jk}, 2+\kappa-k_{jk}, -1\right)}{(2+\kappa-k_{jk})! \, (d_{kj}-\kappa+k_{jk})}$$

In terms of calculating the within and cross-partition communication rates for the $\kappa$-radius partitioning, the probability of remaining within $C_J$ is given by:

$$\hat{A}_{JJ} = 2 \sum_{j \in J} p^*_j w_j \sum_k p^*_k w_k T_{jk \to i \in J} =$$

$$2 \sum_{j \in J} p^*_j w_j \sum_k \sum_{r=0}^{\kappa - r_{jk}} p^*_k w_k \binom{d_{jk}}{r} \left(\frac{1}{2}\right)^{d_{jk}}$$

To compute the cross-class communication rates, one needs to define other equivalence classes in terms of Hamming distances to some set of reference vertices, i.e.

$$\sum_{I \neq J} \hat{A}_{IJ} = 2 \sum_{j \in J} p^*_j w_j \sum_k p^*_k w_k T_{jk \to i \notin J} =$$

$$2 \sum_{j \in J} p^*_j w_j \sum_k \sum_{r=\kappa - r_{jk}+1}^{d_{jk}} p^*_k w_k \binom{d_{jk}}{r} \left(\frac{1}{2}\right)^{d_{jk}}$$

which, as for the sum over values within the $\kappa$ radius, has no closed form but evaluates to a multiple of the Hypergeometric function $H_{21}$:

$$P(d_{0\,i} > \kappa) = \sum_{k=\kappa - k_{jk}+1}^{d_{jk}} \binom{d_{jk}}{k} \left(\frac{1}{2}\right)^{d_{jk}} =$$

$$\frac{2^{-N}(1+d_{jk})!\, H_{21}(1,\, 1+\kappa - k_{jk} - d_{jk},\, m+\kappa - k_{jk},\, -1)}{(2+\kappa - k_{jk})!\,(d_{kj} - \kappa + k_{jk})}$$

Unfortunately, there is no way to collect $p_j w_j$ terms for partitions thus defined and derive a result in terms of the mean fitnesses of each class. The reason for this is that while for uniform recombination schema-based equivalence classes, each value $T_{jk \to i}$ is equal for all k∈K, allowing one to factor out a $\overline{W}_J$. For recombination across Hamming-distance based equivalence classes, $T_{jk \to i}$ will differ according to one's choice of k. This is another reason why there is no closed-form expression for the within versus cross-class communication rates for this partition. Consequently, as far as we can determine the only way to evaluate whether any Hamming distance based partitioning is consistent with Simon-Ando is by numerical computation on a case-by-case basis for each fitness function.

The situation is of course even more problematic for single-point (or multipoint) recombination, in that the offspring classes have to be averaged over all parental types of given Hamming distances d of one another, as the probability of any offspring type depends not only on the Hamming distances but on the lengths separating individual loci.

However, the analysis of uniform recombination suggests one result that almost certainly applies to most recombination-selection systems. Because most modes of recombination allow the production of offspring outside the parental partition at relatively high rates (as long as all possible parental pairings are permitted with equal probability), it implies that in general achieving the conditions for Simon-Ando with recombination in each generation is unlikely. One rather trivial way of achieving a much higher rate of within-partition versus cross-partition communication is to reduce the frequency of recombination events, i.e. as in the scenario investigated above, have probability $\rho$ of sexual reproduction and 1-$\rho$ of selfing. For $\rho \ll 1$, most offspring are identical to their parents and hence by default remain in whatever partition they belong in.

A more interesting model which we leave as an open-ended question is what happens under assortative mating. If instead of random mating we assume that there is some pairing probability function f($p_j, p_k$) that depends on the pairwise Hamming distances between j,k or on the partial Hamming distances between relevant schema-defining loci, it should be obvious that if the parents tend of both be members of the same partition there will be a bias in producing offspring which are members of this partition. This is a more interesting scenario than the facultative recombi-nation model because it allows for a non-trivial internal transmission dynamic within and across partitions, as well as relating conceptually to the origin of species through assortative mating mechanisms (Bush 1982, Kondrashov and Minna 1986).

## Discussion: Future Directions

The question of just how "common" fitness landscapes with transition operators consistent with (3.0) and by necessity (6.1) relative to the entire space of possible fitness landscapes might be can be addressed from a number of directions. A natural approach would be to define a "landscape space" for an n-locus system, i.e. as a Fourier function space $f: V \mapsto \mathbb{R}$,

$$(9.1) \qquad f(v) = \sum_{k=1}^{N} a_i \phi_k(v) + a_0$$

where $\phi_k(x)$ is a rule mapping each k-tuplet subset genotype v onto a real-valued contribution to fitness. Representing each k-tuplet as $\sigma_1 \sigma_2 ... \sigma_k$ where $\sigma_i = \pm 1$, we have a straightforward representation of epistatic interactions as Walsh function components (Weinberger 1992, Stadler 1994, Stadler et al 1998). Given a k-subset of loci $\pi_k$ and summing over all $\binom{n}{k}$ possible k-products:

$$\phi_k = \sum_{x=1}^{\binom{n}{k}} \prod_{i \in \pi_x} \sigma_i$$

Now given a parameter space $\{a_0 ... a_n\}$, we have a well-defined fitness function space in this way, we can ask what proportion of landscapes in this universe have a partitioning which satisfies (6.1). Given the intractable dimensionality of this function space, a more reasonable starting point might be simply to ask which orders of epistatic interaction (i.e. fitness functions

f=$\phi$, given "elementary landscapes" sensu Stadler 1994) tend to produce the desired landscapes.

An alternative method would be to characterize "valley" landscapes in terms of landscape autocorrelation (Stadler 1994). Landscape autocorrelation has been related to orders of epistatic interaction, there is a straightforward relationship between the statistical characterisation and the underlying epistatic rules that we can make use of. Such an approach may be fruitful because the conditions necessary for valley landscapes seem to require strong local fitness function autocorrelation (producing localized neighborhoods of high fitness genotypes) and globally low autocorrelation (leading to separation of the high fitness partitions by valleys).

Work on landscape connectivity by Gavrilets and others (Gavrilets and Gravner 1997) suggests that high levels of connectivity are a generic property of high-dimensional systems while separation by broad valleys requires rather specialized conditions on the fitness functions. So whether decomposable systems prove relevant is probably significant empirically than theoretically. We know that they can be constructed even though most landscapes probably are not of this form, the question remains whether this more restricted class of landscapes is common enough for aggregation of variables methods to be at all useful in analyzing mutation-selection systems.

Therefore, what one is ultimately interested in is not whether random landscapes satisfy Simon-Ando or the "valley" generalization, but whether fitness landscapes in nature (i.e. real genetic systems) tend to have the structure in question. This is a more complicated question to address given the difficulty of measuring fitness in any but the most artificial systems, though a good starting point may be the growing literature on RNA and protein landscapes defined in terms of the polymer's performance of a given catalytic function.

On a final note, we return to an observation we made in the introduction, namely, that aggregation of variables (at least implicitly or conceptually) has found numerous applications in evolutionary biology. An approach in many ways complimentary to ours has been undertaken by a number of groups (Frenkel et al 2000, Watson 2002), where partitioning according to strong versus weak coupling is used to represent epistatic interactions. In their case, the Simon-Ando partitions correspond not to clusters of genotypes but rather to clusters of interacting genes. Whether or not there is between decomposability of epistasis rules and landscape decomposability as defined in this paper remains to be answered.

## ■ Dedication and Acknowledgements

We dedicate this paper to the memory of the late Dr. Herbert Simon (1916-2001), with whom the senior author had the privelege of corresponding with in the last year of his life, and whose original research inspired this project.

We express our gratitutude to Joachim Hermisson for his advice on recombintaion models and for his collaboration on related projects. The authors also thank the following people for their assistance and commentary on both technical and conceptual aspects of this paper: Sean Rice, Junhyong Kim, Robert Dorit, Ashley Carter, Pierre-Jacques Courtois, Claus Wilke, and Richard Watson.

# ■ Figures

**Figure 1**:

Shows a plot of the exact stationary distribution (uniform at 1/32 for each genotype) against the Simon-Ando eigenvector estimate for a 5-locus, 2 allele fitness landscape with every genotype fitness $W(x)=1.0$. The 32 genotypes are arranged in the order of Boolean numbers, i.e. 00000,10000,...11111.The mutation rate per locus is set to 0.01. While the order of magnitude estimates are correct, the Simon-Ando approximation for any particular genotype deviates due to the fact that edge vertices in each partition class are set as "boundaries" in the approximation.

**Figure 2**:

The following plots are for the mutation-selection matrix induced by a bimodal fitness landscape. In a 5 locus, 2 allele system, the genotypes 00000 and 11111 have fitness values set to 1.0. The Hamming distance one genotypes have a fitness of 0.9, while the Hamming distance two neighbors (with respect to either peak) are set to 0.01. Mutation rate is again $\mu=0.01$.

a) This symmetric, bimodal fitness landscape has degenerate leading eigenvectors with identical $\lambda=0.999$ eigenvalues. The normalized degenerate eigenvectors are shown above.

b) The stationary distribution for this bimodal landscape is plotted against togehter with the estimated distribution derived from the eigenvectors of the Simon-Ando partition matrices.

**Figure 3**:

The fitness landscape is the same as the above, except that one peak was set to a lower fitness ($W(11111)=0.9$, $W(00000)=1.0$) value to illustrate directional selection. The corresponding unimo-dal leading eigenvector is closely approximated by Simon-Ando aggregation.

**Figure 4**:

The fitness landscape in this set of computations has two peaks of almost equal fitness: W(00000)=1.0 and W(11111)=0.99. However, the peak with somewhat lower fitness has mutational neighbors with fitness values approximately equal to 0.9 while the higher peak has single-point mutational neighbors with a fitness of about 0.1. The Hamming distance two neighbors with respect to both peaks are again set to 0.01, with $\mu$ set to 0.01. For this parameter range, the stationary distribution's probability density is concentrated about the more "mutationally robust" genotype, i.e. the one with the higher fitness neighbors. While this pattern is qualitatively predicted using the Simon-Ando approximation, it can be seen from the graph that the prediction is rather poor.

# ■ References

# Appendix:

Max Shpak, Department of Ecology and Evolutionary Biology, Yale University. New Haven, CT 06520-8106, USA

Joachim Hermisson, Deptartment of Biology, University of Munich, Luisenstrasse 14, D-80333 Munich, Germany

Peter Stadler, Institute of Bioinformatics, University of Leipzig, Kreuzstrasse 7b, D-04103, Leipzig, Germany

# ■ Generalization of Simon-Ando Decomposability: Fitness Landscape Applications

One of the limits to applicability of Simon-Ando decomposition to mutation-selection systems is the highly constrained type of vertex paritioning which defines the model. The structure of a transition matrix **A** that is consistent with Simon-Ando decomposability has the block diagonal form (3.0), i.e. with block-diagonal submatrices where the off-diagonal elements are all of order $\epsilon \ll 1$.

The reason that a partitioning consistent with (3.0) is so restrictive is that it requires that every vertex be a member of a partition where every element communicates strongly with some of the vertices that are members of the same subset (see closure theorem 6.2). Consequently, a strict Simon-Ando partitioning allows for only as many partitions are there are local optima, thus a two-peaked landscape can only be partitioned into two subsets, at least one of which has a large number of contituent microstates. This does not permit a substantive reduction of state space unless there are a large number of peaks. Furthermore, because the within versus cross-partition communication rates for partitions of arbitrary topology do not generally have closed form solutions, we proposed the use of $\kappa$-ball partitions to simplify the aggregation calculations. As noted above, however, a generic landscape need not be partitionable into $\kappa$-balls of fixed radius. It is necessary to allow partitions of varying radius, in many cases including trivial "partitions" which consist of a single vertex. If many such vertices remain in the complement to the $\kappa$-ball partitioning, it becomes apparent that the original goal of an aggregate representation is only partially satisfied.

Here we investigate the possibility of treating the "remainder" vertices in a landscape which is not fully partitionable as a self-contained aggregate entity in its own right independent of its topology or fitness distribution. What we propose below in preliminary form is a weaker version of Simon-Ando decomposition that would allow one to make partitions around the optima and to have another partition (or any number of partitions) corresponding to the low-fitness genotypes separating the two peaks.

The somewhat less restrictive "block triangular" form discussed in Ando and Franklin (1963) would still impose the same constraints on landscape partitioning, while alternative aggregation of variables techniques (such as the bounded aggregation methods of Courtois 1984, 1989) also don't give representations consistent with the partition classes and their complements used in our class of models. Here we analyze a fitness landscapes with a matrix structure where part of the lattice is partitionable into Simon-Ando components while a substantial complement set of vertices remains which is not a Simon-Ando partition according to definition (6.1).

The classical view of Wrightean landscapes postulates low fitness "valleys" of arbitary width (with any number of steps separating high fitness regions). In this case, with a proper arrangement of rows and columns, where an arbitrary number of diagonal "submatrices" are also of order $\epsilon \ll 1$

$$(8.0\ a) \quad \begin{pmatrix} A_1 & \epsilon & \epsilon & \epsilon & \epsilon \\ \epsilon & \ddots & \epsilon & \epsilon & \epsilon \\ \epsilon & \epsilon & A_I & \epsilon & \epsilon \\ \epsilon & \epsilon & \epsilon & \ddots & \epsilon \\ \epsilon & \epsilon & \epsilon & \epsilon & E \end{pmatrix} = \boldsymbol{A}^* + \epsilon \boldsymbol{C}$$

or, in the special case of no direct (single-step) communication between any two of the partitions I,J, the corresponding matrix has the structure

$$(8.0\,b) \quad \begin{pmatrix} A_1 & 0 & 0 & 0 & \epsilon \\ 0 & \ddots & 0 & 0 & \epsilon \\ 0 & 0 & A_I & 0 & \epsilon \\ 0 & 0 & 0 & \ddots & \epsilon \\ \epsilon & \epsilon & \epsilon & \epsilon & E \end{pmatrix} = \mathbf{A^*} + \epsilon \mathbf{C}$$

where E is a square matrix of order $\epsilon$ values of arbitrary dimension, corresponding to a decomposition $\mathbf{A} = \mathbf{A^*} + \epsilon \mathbf{C}$ where $\mathbf{A^*}$ consists of block diagonal matrices up through the Kth partition and zeros elsewhere.

A transition matrix in such a form would correspond to a scenario where there are K sets of strongly communicating vertices are separated by a single "valley" region of weak communication. Note that the particular matrix arrangement is not significant: one can divide up the valley region such that individual blocks $A_I$ and $A_J$ are separated by near-zero regions, for the purposes of simplifying notation we find it best to cluster the "junk" regions into a single block of size N-$\sum_I n_I$.

Unlike the Simon and Ando partitioning defined in (3.0), such a model places no topological constraints on the position and number of low-fitness vertices, requiring only the existence of high fitness neighborhoods iin various parts of the landscape (a reasonable assumption if fitness values are locally correlated).

Furthermore, the dynamical properties which are of interest in Simon-Ando systems should still be applicable in this general case. Because $\mathbf{A} \simeq A^*$, the some of the results of Theorem (2.2) hold for a dynamical system defined by a matrix of the form (3.0). As before, we should expect there to be short-term quasiequilibria $x_J^*$ associated with the relatively high-valued block partitions $A_J \simeq A_J^*$. Stated formally, we can approximate the short-term dynamics of $x_{j_J}$ according to (3.4), i.e. for t less than some critical valued $T_2$,

$$x_{j_J}(t) \simeq x^*_{j_J}(t) = \lambda_{1_J}^{*t} z^*_{1_J\ j_J} Y_{1_1}(0) + \sum_{i_J=2}^{n_J} \lambda_{i_J}^{*t} z^*_{i_J\ j_J} Y_{i_J}(0)$$

The differences between a system defined by a matrix of form (8.0) and Simon-Ando arise in the dynamics during $T_2 < t$. In Simon-Ando systems, the transition rate between I,J is approximated by $\sum_{i \in I} \sum_{j \in J} x_j A_{ij}$, as the strongest links between I,J are single-step transitions between the partitions. In contrast, for (8.0) the sum across all paths through from a vertex in I to one in J through **E** (the N-$\sum_I n_I$ lowermost block of order $\epsilon$ terms) give transition rates of the same order (or higher) than the direct communication between I and J. Indeed, one can construct models where the only connections between certain peaks and plateaus is through multistep mutational walks through the valleys. Therefore, the approximation of $\hat{A}$ by (3.13) will not be applicable to this generalized "valley" model.

In terms of the spectral decomposition of **A**, the contributions of the last, order $\epsilon$ diagonal block **E** come in through the last term in the expansion below. The starting point for the **E** block is at $\sum_I n_I = n_E$, with the index endpoint at N as before, with the high probability density partitions $C_1 ... C_K$,

$$(8.1) \quad x_{j_J}(t) = \lambda_{1_1}^t z_{1_1\ j_J} Y_{1_1}(0) + \lambda_{1_J}^t z_{1_J\ j_J} Y_{1_1}(0) +$$

$$\sum_{I \neq J, I=2}^{K} \lambda_{1_I}^t z_{1_I\ j_J} Y_{1_I}(0) + \sum_{i_J=2}^{n_J} \lambda_{i_J}^t z_{i_J\ j_J} Y_{i_J}(0) +$$

$$\sum_{I \neq J, I=1}^{K} \sum_{i_I=2}^{n_I} \lambda_{i_I}^t z_{i_I\ j_J} Y_{i_I}(0) + \sum_{i_E=n_E}^{N} \lambda_{i_E}^t z_{i_E\ j_J} Y_{i_J}(0)$$

Generally, there is no hierarchical structure in **E**, so that one cannot generally decompose the eigenspace associated with this block in a way that the leading eigendirections within the high-valued blocks $A_I$ tend to separate out (i.e. as the second versus the third terms in the above expression).

Theorem (2.2) applies to the distribution components associated with the high-valued partitions. However, for the submatrix **E** there is no necessary difference in communication rate for vertices within **E** and between **E** and the block partitions (indeed, the communication rate of $C_E$ with other partitions will tend to be higher, at least at the edges). Therefore, because the decomposition $\mathbf{A}=\mathbf{A}^*+\epsilon\mathbf{C}$ still holds for this system, we again have a ranking of eigenvalues $\lambda_1...\lambda_n$ where the first K eigenvectors are approximated by the eigenvectors associated with $\lambda^*_{1_1}...\lambda^*_{1_K}$ (the leading eigenvalues for each of the K blocks in $A^*$) while eigenvectors associated with the off-diagonal (cross-parition terms) and with the block **E** are of order $\xi$ (as defined in Theorem 2.1).

Following the notation used in the (3.5) spectral decomposition,

$$(8.2) \quad x_{j_J}(t) = \xi \lambda_{1_1}^t u_{1_1 j_J} Y_{1_1}(0) + \lambda_{1_J}^t z_{1_J j_J} Y_{1_1}(0) +$$

$$\xi \sum_{I \neq J, I=2}^{m} \lambda_{1_I}^t u_{1_I j_J} Y_{1_I}(0) + \sum_{i_J=2}^{n_J} \lambda_{i_J}^t z_{i_J j_J} Y_{i_J}(0) +$$

$$\xi \sum_{I \neq j, I=1}^{m} \sum_{i_I=2}^{n_I} \lambda_{i_I}^t u_{i_I j_J} Y_{i_I}(0) + \xi \sum_{i_E=n_E}^{N} \lambda_{1_I}^t u_{1_I j_J} Y_{1_I}(0)$$

which gives us, in the place of (3.7b),

$$(8.2b) \quad x_{j_J}(t) = \xi S_j^{(1)} + S_j^{(2)} + \xi S_j^{(3)} + S_j^{(4)} + \xi S_j^{(5)} + \xi S_j^{(6)}$$

As for Simon-Ando systems, Theorem (2.2a) holds. Equation (3.8) applies to (8.2) (thus defining the interval $t<T_1$ at which the non-leading eigendirections of $A_I^*$ are significant), while the only modification of (3.9) comes from the $S^{(6)}$ term in (8.2b), i.e.

$$(8.3) \qquad \frac{\xi\left(S_j^{(1)} + S_j^{(3)} + S_j^{(5)} + S_j^{(6)}\right)}{S_j^{(2)} + S_j^{(4)}} < \eta_1$$

For $T_1<t$ cross-partitions communication (including the contribution of eigendirections associated with valley $\mathbf{E}$ in term $S^{(6)}$) begin to dominate the system dynamics. Similarly, Theorem 2.2B applies in the valley system as there exists a $T_2$ such that for $T_2<t$ the leading eigendirections of each large-valued partition begin to dominate

$$(8.4) \qquad \frac{S_j^{(4)} + \xi S_j^{(5)} + \xi S_j^{(6)}}{\xi S_j^{(1)} + S_j^{(2)} + \xi S_j^{(3)}} < \eta_2$$

and furthermore, there exists $T_2<T_3$ so that for $T_3<t$ only the leading (globally dominant) eigendirection $S^{(1)}$ is significant:

$$(8.5) \qquad \frac{S_j^{(2)} + \xi S_j^{(3)} + S_j^{(4)} + \xi S_j^{(5)} + S_j^{(6)}}{\xi S_j^{(1)}} < \eta_3$$

It follows from the above that the "valley" system exhibits the fast-slow behavior analogous to that of Simon-Ando dynamics, which in turn implies a quasi-independence (decomposability) of the system with respect to the high-valued partitions $C_I$ over the short term. Since for times $T_1<t<T_2$ the state distribution x(t) is dominated by $v_1(1)...v_K(1)$, we can use the approximation of x(t) by (3.12), i.e.

$$x_{i_I} \simeq v^*{}_{i_I}(1_I) \sum_{i \in I} x_{i_I}$$

Within the time interval $T_1 < t < T_2$, treating the partition frequencies $X_I = \sum_{i \in I} x_{i_I}$ as aggregate variables should give close approximations. The problem arises in computing the dynamics at $T_2 < t$ (when cross partition communication becomes significant) because one needs an estimate of transition rates between partitions $C_I$ and $C_J$. In contrast to Simon-Ando, the weighted sums (3.12) across off diagonal submatrices does not give a satisfactory estimate of cross-partition transition rates.

While in Simon-Ando systems complete partitionability guarantees that indirect paths connecting $C_I$ and $C_J$ were insignificant because any such path necessarily passes through another partition $C'$ that would act as a short-term sink, this is not the case for transition operators of the form (8.0). The valley $C_E$ does not act as a quasi-absorbing state, hence paths through the valley are significant, and, in the case of a matrix of the form (8.0b), completely determine the system dynamics. Consequently, in computing the transition rates between any two high-valued partitions $C_J$ and $C_I$, what we are often interested in is an estimate of the passage rate from $C_J$ to $C_E$ and from $C_E$ to $C_J$.

Using the aggregation methods from Simon-Ando systems, we know that the direct transition rate between any two partitions can computed using (3.13). For $T_2 < t < T_3$, we can apply the same formula to computing the transition rate from any partition $C_J$ to $C_E$, i.e.

$$\hat{A}_{EJ} = \frac{1}{X_J} \sum_{i \in E} \sum_{j \in J} A_{ij} x_{j_J} \simeq \sum_{i \in E} \sum_{j \in J} A_{ij} v^*{}_{j_J}(1_J)$$

However, these approximation methods are not applicable to estimating $\hat{A}_{IE}$. While the first part of the expression still holds, the estimate of $x_{i_E}$ by block **E** leading eigenvector

$v^*_{i_E}(1_E)$ is not valid. The approximation of the within-partition distributions by the leading eigenvector does not apply because the order $\epsilon$ coefficients of **E** do not guarantee a strong time decoupling between within $C_E$ dynamics and global dynamics. Therefore, there is no approximation for $\hat{A}_{EJ}$ that can be made without computing the distribution $x_{i_E}(t)$ exactly,

$$\hat{A}_{IE} = \frac{1}{X_J} \sum_{i \in I} \sum_{j \in E} A_{ij} x_{j_E}$$

In view of this constraint, we see two ways in which to deal with the lack of aggregability on component $C_E$.

The most general treatment is to regard $C_E$ as a black box and simply compute expected transition rates (or passage times) between high valued partitions $C_I$, $C_J$ through all possible paths. An alternative is to find limiting cases where the distribution $x_E$ can be reasonably approximated without exhaustive microstate (vertex) frequency calculations, in other words finding fitness functions where the within-vallye distribution $x_E$ could be reasonably treated as a macrostate variable like the Simon-Ando type partitions. Below we outline these possible approaches to aggregation in the valley system:

The exact quantity required in making an estimate of cross partition transition rates scales in inverse proportion to the expected first passage time between any vertex $i \in C_I$ and $j \in C_j$. The passage time to some vertex x can be computed using the matrix $\tilde{A}_x$, which has the coefficients (with columns renormalized):

$$\tilde{A}_{x,ij} = 0 \text{ for } j = x, i \neq x$$
$$A_{ij} \text{ elsewhere}$$

Which makes x an absorbing state and guarantees that the passage time to x is non-recurrent. If $\mathbf{A}$ and $\tilde{\mathbf{A}}$ are stochastic matrices, the mean first passage time for $\tilde{\mathbf{A}}_x$ is computed by

$$(8.5\,a) \quad E(t_{ij}) = \sum_{k=0}^{\infty} k \left( \tilde{A}_{ij}^k - \tilde{A}_{ij}^{k-1} \right) = 1 + \sum_{k=0}^{\infty} \left( 1 - \tilde{A}_{ij}^k \right)$$

The difference $\tilde{A}_{ij}^k - \tilde{A}_{ij}^{k-1}$ is used because $\tilde{A}^k$ itself measures the probability that a transition from y to x has occured by time within k steps, not necessarily at the kth step. However, this is not applicable to mutation-selection matrices because they are not row/column stochastic, i.e. the transition rates reflect the differences in fitness values between the parent states.

For any transition matrix $\mathbf{A}$, we must rewrite the coefficients in terms of probabilities in order to compute expected passage times. We define the quantity

$$P^{(k)}{}_{ij} = \frac{A^{(k)}{}_{ij}}{\sum_{i=1}^{n} A^{(k)}{}_{ij}} = \frac{A^{(k)}{}_{ij}}{\varsigma_j^{(k)}}$$

which for k=1 is simply the mutation rate $\mu_{ij}$, while for iterations k>1, the term can be interpreted as a generalised multigeneration mutation rate (i.e. measuring the probability that parent j produces progeny i in k steps). In turn, the sum term $\varsigma_j^{(k)}$ is $W_i$ for k=1 and can be interpretted as a generalised fitness term which measures the effect of ancestral fitness for a lineage over k generations.

The probability that an ancestor in state j has a descendent i in k generations is

$$1 - \left( 1 - p_{ij}^{(k)} \right)^{\varsigma_j^{(k)}}$$

From this we derive the generalised expected first passage time

$$(8.5b) \quad E(t_{ij}) = \sum_{k=0}^{\infty} k\left((1-p_{ij}^{(k)})^{\zeta_j^k} - (1-p_{ij}^{(k-1)})^{\zeta_j^{k-1}}\right) = 1 + \sum_{k=0}^{\infty} (1-p_{ij}^{(k)})^{\zeta_j^k}$$

It can be seen that in the limiting case of a stochastic matrix (where $A_{ij}=p^k{}_{ij}$ for all k), (8.5b) is identical to (8.5a) since $\zeta=1$. We note that measures of generalized lineage fitness $\zeta^k$ across k-generations is conceptually and mathematically equivalent to the clonal fitness measures in Hermisson et al (2002).

From (8.5), computing the expected first passage time between two partitions I,J is straightforward given (8.5),

$$(8.6) \quad E(t_{I,J}) = \sum_{i \in I} \sum_{j \in J} x_j E(t_{ij})$$

is the exact value of expected passage times between partitions $C_J$ and $C_I$.

The construction of (8.0) is such that the shortest paths between any two partitions $C_I$ and $C_J$ will generally be through $C_E$ rather than through any other partition. Therefore, any path connecting vertices $i \in C_I$ and $j \in C_J$ via another high-valued partition C' can be ignored in a first order approximation. The transition time (8.6) will be determined largely by direct communication terms (off-diagonal submatrices matrices $A_{IJ}$, $A_{JI}$, if they are nonzero contra 8.0b) and by paths through $C_E$.

We approximate the transition rates by writing **A'** defined so that (with appropriate column normalization)

$$(8.7) \quad A'_{ij} = A_{ij} \text{ for } i, j \in C_I, C_J, C_E$$
$$\qquad \qquad 0 \text{ otherwise}$$

which by removes all coefficients except those associated with the two partitions of interest and the valley (in other words, both coefficients i and j must correspond to vertices which belong to $C_I, C_J, C_E$). To compute expected first passage times between I and J, we use (8.5-6) with $\tilde{A}'$ in the place of $\tilde{A}$. For a transition matrix of the form (8.0b), the passage time is:

(8.8)  $E(t_{IJ}) = E(t_{IE}) + E(t_{EJ})$

From estimates of first passage times we can come up with measures of transition rates per iteration as $\hat{A}_{IJ} = \frac{1}{E(t_{IJ})}$. This allows one to compute the cross-partition communication rate on the same time scale as transitions between individual vertices. Given the aggregate transition matrix $\hat{A}$, we compute $X_I(t)$ and use the coefficients to estimate those coefficients according to (3.12), i.e. $x_{i_I}(t) \simeq X_I(t) v_{i_I}(t)$ for those $x_i(t) \notin C_E$.

Unfortunately, when a substantial portion of the state space of a large dimensional system is occupied by the low valued coefficients in **E**, little is gained from the standpoint of computational efficiency by an aggregation of variables approach (on the other hand, whatever conceptual insights are gained from an aggregation of genotypes into macrostate variables still apply here due to the quasi-independence of the partitions).

However, if we assume that coefficient values of **E** correspond to some limiting case, then we can treat the valley $C_E$ as an aggregate variable in its own right, thereby significantly reducing the computational complexity. We propose two limiting cases which could give reduced representations of $x_{j_E}$. One is a scenario where all coefficients corresponding to vertices $x \in C_E$ assume the same order of magnitude in value. Thus the "valley" becomes a flat plain separating the peaks, with no two paths of the same length through the valley dominating (rather, shortest paths determine the best estimates of effective transition rates). We illustrate this schematically in Figure 1.

Another limiting case would be the following scenario: while every element of **E** has substantially lower fitness than any coefficients in $C_1...C_K$, there is some subset of connected elements in $C_E$ which have significantly higher fitness than the others, creating an effective high(er) fitness "ridge" in the valley which connects at least two of the high valued partitions (see Fig. 2). In such a case, the transition rate between two partitions is closely approximated by the rate of movement along the ridge (whereby a single path almost completely determines the expected passage time $t_{IJ}$).

The importance of these two limiting cases is that both allow one to estimate passage times without information about the probability density at each valley vertex $x_{i_E}$, allowing one to (effectively) treat $x_E$ itself as a macrostate variable. In the case of the ridge, most of the probability density in $C_E$ would be on the ridge itself, while for the planar valley, most of the probability density in $C_E$ would be in the vicinity of the neighboring peaks with a nearly uniform distribution elsewhere in $C_E$.

If every genotype in the valley region has (approximately) the same fitness $w_E$, the expected first passage time from one partition to another can be computed using only the adjacency matrix **D** where $D_{ij}=1$ when $A_{ij}>0$ and is 0 elsewhere. If we are interested in the traversal time between any two partition, then we compute the adjacency matrix for the "reduced" operator **A'**, as defined in (8.6). The number of paths (including loops) of length k from any vertex pair j to i is $D_{ij}^k$.

This suggests an immediate estimate for the probability of transitition between any vertex $j \in C_J$ and $i \in C_I$ within time t. Under the assumption that all valley fitness values are equal, we apply the standard binomial distribution model for a homogeneous random walk of t steps.

$$(8.9) \quad \Pr(j \to i, t) = w_j p_j C \sum_{k=0}^{t} D_{ij}^k w_E^t \binom{t}{k} \mu^k (1-\mu)^{t-k}$$

with C the normalization constant

$$C \sum_{t=0}^{\infty} \sum_{k=0}^{t} D_{ij}^k w_E^t \binom{t}{k} \mu^k (1-\mu)^{t-k} = 1$$

so the expected transition time is given by

$$(8.10) \quad E(t_{ij}) = w_j p_j \sum_{t=0}^{\infty} t C \sum_{k=0}^{t} D_{ij}^k w_E^t \binom{t}{k} \mu^k (1-\mu)^{t-k}$$

We estimate the cross partition transition rate $\hat{A}_{IJ}$ by applying (8.6) to the above and taking the reciprocal as in the general case.

For many landscapes, (8.10) can be closely approximated by summing over only the k-values and time intervals for the first few shortest paths, because for $w_E \ll 1$ the contribution of the longer paths becomes insignificant as $w_E^t \to 0$ (for large values of t,k such that k≤t). It should also be noted that (8.9) and (8.10) are only approximate because there is an implicit assumption that paths with loops only travel through fitness $w_E$ valley genotypes whereas in reality a certain portion of paths pass back through the high-fitness partition sets.

If there is a symmetry to the partition classes $C_1 \ldots C_K$ and to $C_E$, further simplification is possible. Consider a landscape where the classes are R-balls centered about origins x, calculating path length and number is straightforward. Consider two partitions $C_1$ and $C_2$, centered about origins $x_1$, $x_2$ and with radii $R_1$, $R_2$ respectively. The distance between the origins is $H(x_1, x_2) = d_{12}$. It can be seen that the shortest path(s) between the partitions (whenever $R_1 + R_2 < d_{12}$) is $d_C = d_{12} - R_1 - R_2$. We denote the next shortest paths as $d_{12}+1, d_{12}+2, \ldots$ respec-tively.

We also require that all edge vertices on the equivalence classes have the same fitness values and (consequently) approximately the same frequencies, so that all paths of the same length are equivalent and do not have to be weighted by the fitness and frequency of the various starting points in $C_J$ (in other words, the fitness values and initial frequencies of the starting points can be factored out as irrelevant scaling terms).

Let $N_k(I,J)$ denote the number of loop-free paths of length k between partitions $C_I$ and $C_J$. Because we are counting loop-free paths of fixed length (as opposed to generalized k-step paths), so the number of steps in the "forward" versus the "reverse" directions have to be taken into account. For a k-path, the constraint on forward and backsteps is k=F-B (each occuring with mutation probability $\mu/2$) while the remaining t-F-B no-step iterations occur with probability $1-\mu$. The probability of following the loop free path of lenght k from one endpoint to the next over t iterations is given by the multinomial:

$$(8.11) \quad M(k,t) = \frac{t!}{F!(F-k)!(t-k-2F)!} w_E^t \left(\frac{\mu}{2}\right)^{2F-k} (1-2\mu)^{t+k-2F}$$

where the coefficient of 2F-k comes from the product $\left(\frac{\mu}{2}\right)^F \left(\frac{\mu}{2}\right)^{F-k}$

Taking into account all k-paths separating I and J, the rate of cross-partition traversal over time t is approximated by

$$(8.12) \quad P_{IJ} = \sum_{k=0}^{\infty} \sum_{t=k}^{\infty} N_k(I,J) M(k,t)$$

With proper scaling and normalization, (8.12) gives an estimate of $\hat{A}_{IJ}$. The same caveats which applied to (8.9-10) hold in this case too, because the $w_E^t$ factor is made with the assumption that even paths with loops usually stay within the valley.

When there is a ridge in the valley, i.e. such that most or all of the other valley vertices have coefficients of order $\epsilon$ while ridge vertices have associated coefficients (fitness values) of the order $\alpha$ where $\epsilon<<\alpha<<1$, we need only compute the transition rate across a single path for a first-order estimate of cross-partition communication. Considering the simplest case, one where there is a single loop and branch free ridge of length k connecting two vertices in the partition, the effective transition rate will be almost entirely determined by the speed at which this path is traversed.

This quantity can be computed using (8.11), the probability of traversing a k-path over t time steps. The cross-partition passage time is simply the expected time to cross the k-path weighted by the frequency of the vertex $r \in C_J$ which connects to the ridge:

$$(8.13) \quad E(t_{IJ}) \simeq \sum_{t=0}^{\infty} t x_r M(k, t)$$

(where the frequency $x_r \simeq v^*_{J_r}$) with the appropriate per-iteration transition rate the reciprocal of this quantity.

For most fitness landscapes, it should be apparent that the computations involved in making the aggregate approximations above are in many cases more costly than computing the dynamics of the entire system. As such, the methods are computationally useful only in those special cases where the transition rate is completely dominated by a single path. However, a generalized representation of near-aggregability in fast-slow dynamical systems is conceptually useful in that it allows one to identify dynamically coherent classes of variables (genotypes) in systems which cannot be described according to Simon and Ando's (1961) models.

Another interesting perspective which comes out of computing the transition probabilities between peaks across k intervals is the concept of "lineage fitness." The transition matrix $A_{ij}^k$ can be interpreted as the expected number of type i descendants of genotype j after k generations, while $\sum_{ji} A_{ij}^k$ represening the expected number of descendants (irrespective of current genotype) of j after k generations. This interpretation was used in deriving the ancestor distriubtion in Hermisson et al (2002).

What relationship (if any) exists between interpretations of lineage fitness and dynamical decomposability remains an open question. It may prove to be the case that the aggregate transition rates and fitnesses of each class of a (generalized) Simon-Ando partitioning correspond to the effective fitnesses and transition probabilities associated with the locally optimal genotypes and their kth generation descendents.

Simon, H.A. and A. Ando (1961). Aggregation of variables in dynamic systems. Econometrica: 29:111-138

## ■ Concluding Remarks

Our analysis of mutation-selection systems suggests that system decomposability related to fast-slow dynamics (Simon-Ando) is only one of the possible forms of dynamical decomposability and aggregation that lead to identifiable state variables above the level of the genotype.

While Simon-Ando decomposability is fairly easily induced on a fitness landscape with localized peaks and their neighborhoods, a number of evolutionarily interesting examples of decomposability do not seem to fall under the class of systems which are separable due to the existence of local quasiequilibria. The most obvious example of course are fitness landscapes where the symmetry properties allow for an exact aggregation of variables over all time scales (i.e. the equitable partitions and their generalization described in the previous chapter). In that case, the macrostate variables are dynamically sufficient due to the fact that their components have identical fitness values and equivalent mutational neighborhoods with other partition classes rather than strong within-class couplings. Indeed, members of the equitable partition classes did not communicate with one another at all under single-point mutation.

Even on fitness landscapes which induce Simon-Ando type dynamics under sufficiently low mutation rates, potentially interesting evolutionary processes often occur in the parameter range where the Simon-Ando approximation starts to break down. For example, we analyzed the case of a landscape where an isolated fitness "peak" competed against a slightly lower but genetically canalized fitness "plateau." When mutation rates were very low and the Simon-Ando approxi-mation gave very good eigenvector estimates, the equilibrium distribution was concentrated at the highest peak irrespective of the fitness values of mutational neighborhoods.

For higher mutation rates, the more interesting result of the lower fitness plateau dominating the distribution occured. This situation can be interpreted as one where selection effectively acts on genetic lineages (i.e. local peaks and their neighborhoods over several generations) rather than simply selecting for the most fit genotype. Yet the mutation rates at which selection favors the canalized (but less fit) genotype are those where the Simon-Ando approximation is quite poor. This suggests that if indeed there is a level of selection above the genotype level responsible for favoring the canalized genotype, this higher unit does not correspond to a Simon-Ando partition class. As with equitable partitioning, the higher level entity is probably defined by something other than strong within-class cohesion.

The form of dynamical decomposability which has probably received the most attention in the recent literature is the identification of evolutionary characters (Wagner and Laubichler 2000, Wagner and Stadler 2003). Given a genotype or phenotype with specified transmission and selection operators, it is asked whether there are subsets of the genotype (e.g. genes or linkage groups) or phenotype (e.g. morphological characters) whose frequencies are dynamically sufficient state variables. For example, in the special case of linkage equilibrium and multiplicative (or additive) fitness effects, the allele frequency at each locus proves to be a dynamically sufficient state variable in that one can describe the change in allele frequencies at each gene without reference to the allele frequencies at other loci.

Wagner and Laubichler (2000) have generalized the concept of linkage equilibrium so that it could be made without reference to a priori defined entities such as genes in any evolutionary system. The importance of their generalization is that it makes explicit the fact that any character decomposition actually involves an aggregation of variables into equivalence classes defined by a common character state. For example, in a three locus, two allele genotype, the equivalence class *1* is defined as the set of all genotypes having the allelic state 1 at the second locus. To compute the character state frequencies, one sums across the frequencies of constituent genotypes, just as one averages over constituent genotype fitnesses to compute the effective (marginal) fitness of the character.

We note that this aggregation of variables has no (necessary) connection to either equitable partitioning or Simon-Ando dynamical decomposability. While it is possible to construct a partition based on schema (character) equivalence classes which exhibit Simon-Ando dynamics, a character-based aggregation of variables is not a consequence of fast-slow dynamics as it holds across all time scales as long as the generalized linkage equilibrium and multiplicative fitness conditions are satisfied. Nor is there any requirement that the fitness functions and mutational neighborhoods of equivalence classes have the symmetry properties of equitable partitions. As such, this is a completely separate class of aggregation models, contrary to some of our previous intuition about the "universality" of Simon-Ando or generalized equitable partitions.

The case of aggregation into character state equivalence classes has other interesting conceptual implications. The representation in Wagner and Laubichler (2000) makes explicit the fact that character state decomposition is actually a form of aggregation of variables. The standard interpretation of a system under linkage equilibrium with additive fitness is that the gene (or "character") is the unit of evolution and that the description of its state dynamics corresponds to a reduction of state variables to a level below the genotype. In fact, gene or character-level description is only possible because the higher-level aggregate variable (equivalence class) is dynamically sufficient. In other words, far from being a lower-level description, the case of "genic" or "character" selec-tion corresponds to the identification of classes of variables above the level of the genotype or phenotype. Specifically, one derives the dynamics of lower-level variables (i.e. genes) from the higher-level aggregate (frequencies of genotype equivalence classes); it is a conceptual error to say that genotype dynamics are "derived" from allele frequency dynamics in any meaningful sense.

The view of genotype frequency dynamics being "constructed" from gene frequency dynamics below implies that allele (or character) frequencies are the fundamental microstates of the evolutionary process, of which genotype frequency dynamics are a macrostate or "epiphenomenon" (much like classical thermodynamics is the study of macrostates derivable from statistical mechanics). In fact, there is in general no such description for allele frequencies, and when there are, the "allelic dynamics" are actually the change in frequency of higher-order equivalence classes. In some real sense, "genic selection" is actually a higher level selection acting on macrostates defined by some fixed character or allelic identity.

At least in the broad sense, aggregation of variables has proven to be an effective representation for biological state variables above the level of the specified microstate. However, there appear to be several classes of aggregation operators which are structurally independent of one another and therefore aren't necessarily characterized by any common properties apart from the fact that an aggregate variable description exists.

The intuitively appealing view that decomposability and aggregation in biological systems are invariably a result of the type of dynamics described by Simon and Ando (or at least some generalization thereof which preserves the strong-weak interactions within and between partition classes, respectively) appears to be false. At the very least, the importance of Simon-Ando dynamics on fitness landscapes appears to be limited, as many of the dynamical behaviors and emergent properties of potential biological interest seem to be due to symmetries in equivalence classes defined by completely different criteria. It remains to be seen whether Simon-Ando dynamics are important in some of the processes suggested by H. Simon himself (2000, unpublished) such as in metabolic pathways, cell-cell interactions, or gene networks, though even there it is quite likely that the biologically relevant aggregate variables are defined by equivalence classes determined by any number of different symmetry properties.